# Using Machine Learning and Data Mining to Leverage Community Knowledge for the Engineering of Stable Metal-Organic Frameworks

Aditya Nandy[1,2], Chenru Duan[1,2], and Heather J. Kulik[1,]*

[1]Department of Chemical Engineering, Massachusetts Institute of Technology, Cambridge, MA, 02139
[2]Department of Chemistry, Massachusetts Institute of Technology, Cambridge, MA, 02139

AUTHOR INFORMATION

**Corresponding Author**

*email: hjkulik@mit.edu, phone: 617-253-4584

ABSTRACT: Although the tailored metal active sites and porous architectures of MOFs hold great promise for engineering challenges ranging from gas separations to catalysis, a lack of understanding of how to improve their stability limits their use in practice. To overcome this limitation, we extract thousands of published reports of the key aspects of MOF stability necessary for their practical application: the ability to withstand high temperatures without degrading and the capacity to be activated by removal of solvent molecules. From nearly 4,000 manuscripts, we use natural language processing and automated image analysis to obtain over 2,000 solvent-removal stability measures and 3,000 thermal degradation temperatures. We analyze the relationships between stability properties and the chemical and geometric structures in this set to identify limits of prior heuristics derived from smaller sets of MOFs. By training predictive machine learning (ML, i.e., Gaussian process and artificial neural network) models to encode the structure–property relationships with graph- and pore-structure-based representations, we are able to make predictions of stability orders of magnitude faster than conventional physics-based modeling or experiment. Interpretation of important features in ML models provides insights that we use to identify strategies to engineer increased stability into typically unstable $3d$-containing MOFs that are frequently targeted for catalytic applications. We expect our approach to accelerate the time to discovery of stable, practical MOF materials for a wide range of applications.

*Introduction*

Metal-organic frameworks (MOFs) have the potential to revolutionize catalysis[1] and functional materials[2] design due to their reticular nature and well-defined, isolated metal sites.[3-4] To meet numerous design criteria, an exhaustive search of MOFs must be carried out for each potential application, beyond the tens of MOFs that an experimentalist could synthesize.[5-6] The modularity of inorganic secondary building units (SBUs) and linkers that are the building blocks of MOFs have motivated the search for new materials via virtual high throughput screening (VHTS).[4, 7-12] Despite challenges in MOF synthesis[13-14] and post-synthetic modification[15], intense synthesis effort has led to rapid growth[16] in the number of reports of experimentally realized MOFs. MOFs with varying pore size, metals, and linker chemistry have demonstrated desirable properties for storage[17-18], separations[7, 19], sensing[20-21], conductivity[22-24], and catalysis[25-29]. Most MOFs are formed with solvent present in their pores that must be removed for such applications. Although new methods for solvent removal have been developed to activate these materials[30-31], many crystalline MOFs collapse upon activation[14, 32-33], rendering them unusable[22].

MOFs must also retain their porosity and structural integrity even at elevated temperatures relevant for practical application (e.g., catalysis).[34-38] Due to these requirements, VHTS efforts to design new, experimentally realizable MOFs rely heavily on expert intuition for selecting candidates for synthesis after lead compounds are identified computationally.[4, 8] While VHTS can be used to identify new materials, the molecular mechanics models that are tractable for screening MOFs have failed thus far to predict the most experimentally relevant aspects of stability (e.g., activation).[12, 39-41] Because stability measures are difficult to predict from even first-principles methods (e.g., with direct simulation of electronic and structural properties such as metal-linker bond lengths[42]), heuristics such as pore size[43] or hard-soft acid base[44] (HSAB) theory (i.e., applied to the SBU) are instead frequently used to predict MOF stability. However, numerous exceptions exist to such rules, indicating that predictions of stability based on heuristics alone will fail.[43, 45-47] Indeed, stability (e.g., with respect to activation) has proven to be challenging to predict[48], and rules derived from observations on small sets of synthesized materials (e.g., for thermal stability) cannot be extrapolated to new materials.[49]

While experimentally validated measures of stability are thus likely preferable to heuristics or computation, a chief limitation is obtaining them in large numbers. It stands to reason that the concerted effort of thousands of researchers worldwide represents a relatively untapped source of knowledge regarding the factors governing MOF stability. Natural language processing[50] (NLP) combined with machine learning (ML) has been used in a similar way to identify optimal synthesis conditions for inorganic materials (e.g. oxides[51], perovskites[52], and zeolites[53]) and for quantifying the role of privileged molecules (i.e., organic structure directing agents) in controlling synthesized zeolite topology by extracting knowledge from published literature.[54] In comparison to these materials, a lack of systematic naming[55-56] and data reporting[57] has limited the scope of similar approaches applied to the design of new MOF materials.[58-59] Where NLP has been applied to MOFs, it has been limited to properties (e.g., surface area) that can readily be obtained from simulation at low computational cost.[57, 60]

Existing curated data sets aimed at addressing stability prediction are based either on simulations[12] or smaller sets of experiments (i.e., from a single data source).[49, 61] To build truly predictive models of MOF stability that can be applied to new materials, it is desirable to compile as broad a data set as possible. In this work, we leverage the extant experimental literature to construct data sets from reports of MOF's thermal stability and stability with respect to activation (i.e., solvent removal). By exploiting experimental reports associated with well-defined structures, we are able to develop unprecedented maps of the relationships between stability and MOF metal, linker, and connectivity variations over thousands of MOFs. We train ML models to make predictions on new MOFs and reveal design principles for stability, enabling us to demonstrate examples where we can now re-engineer unstable materials into stable ones.

*Dataset Curation*

We started from the 10,143 crystal structures in the all solvent removed (ASR) portion of the 2019 Computation-ready, experimental metal-organic framework (CoRE) MOF database v1.1.2[62]. From this initial set, RACs and geometric descriptors could be computed for 9,597 MOFs (Supporting Information Table S1). Nearly all (9,202 of 9,597) of these MOFs are post-processed structures obtained from the Cambridge Structural Database (CSD) and identified with a unique CSD refcode, 8,809 of which are associated with a digital object identifier (DOI) (Supporting Information). As a manuscript may contain multiple structures, the 8,809 CSD refcodes correspond to 5,152 unique DOIs, 3,809 of which could be downloaded in HTML or XML format (Supporting Information Tables S2–S3). In total, we extracted 7,004 featurizable structures from 3,809 papers for natural language processing (NLP) property extraction (Figure 1).

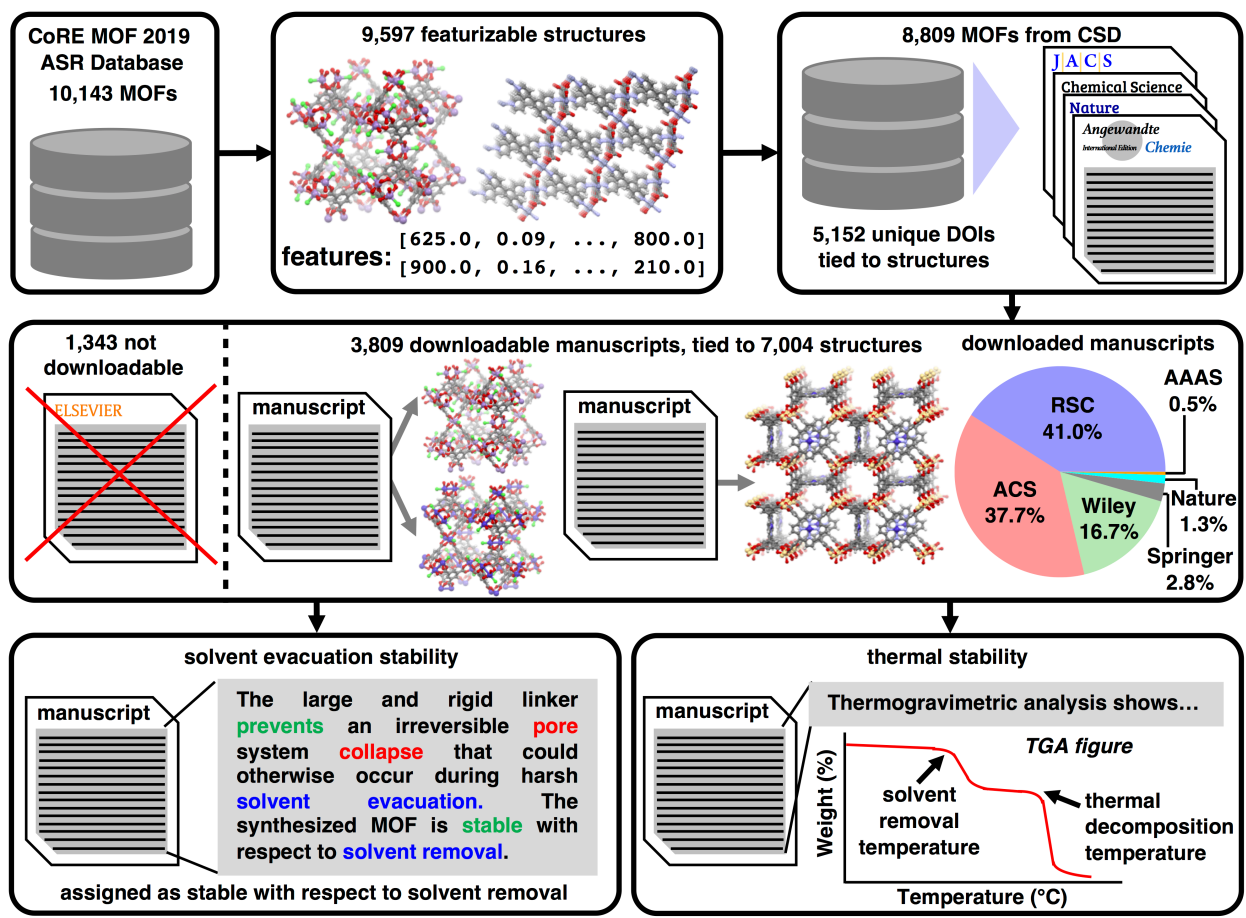

**Figure 1.** Schematic for curation of the datasets used in this work: obtaining MOFs from published work, filtering for reasonable MOF structures that can be featurized, downloading the manuscripts tied to the structures, and text mining the manuscripts for solvent-removal stability or identification of thermogravimetric analysis data.

Next, we parsed articles by sections and tokenized these sections into sentences for further analysis, excluding identifiable (i.e., with headings) introduction sections. If a manuscript had no introduction section heading, we ignored keyword matches in sentences in the first 40% of the manuscript, a cutoff that was chosen by trial and error (Supporting Information Figure S1). For stability upon solvent removal, we pattern-matched (i.e., using regular expressions) sentences that contained keywords simultaneously pertaining to common MOF solvents or to the process of activation (e.g. "dehydration") and to structural integrity (Supporting Information Tables S4–S5). This pattern matching alone cannot be used to determine whether or not a material is stable due to the frequent presence of complex phrases (e.g. "no loss of crystallinity"), negations (e.g. "not stable"), or varied language (e.g. "diminished stability"). After identifying sentences by the automated regular expression check, we analyzed all pattern-matched sentences with syntactic dependency parsing using NLP tools[63] to automatically quantify relationships between words in these phrases. The output of this analysis is a pairwise mapping of related words in the sentence, which simplifies identification of affirmative and negative statements (Supporting Information Figure S2). This approach is essential to distinguish affirmative mentions of stability (e.g. "this material is stable"), negations (e.g. "this material is not stable"), double negations (e.g. "this material is not unstable"), and complex sentence structure (e.g. "none of the MOFs are stable"). We also eliminated sentences about air- and water-stability that were not relevant to solvent-removal stability (Supporting Information Table S6). Using the output of this analysis, we assigned each sentence a binary flag of "stable" or "unstable" with respect to solvent removal. Because most manuscripts discuss more than one MOF structure, we assign the solvent stability label to manuscripts in which all solvent-evacuation-related sentences agree, and we do not automatically assign labels to the structures from manuscripts where conflicting sentiment is present due to the challenge of automated disambiguation of specific MOF stability in these cases (Supporting Information Figure S3).

To quantify thermal stability, we used pattern matching to identify sentences mentioning thermogravimetric analysis (TGA), which suggest that a TGA trace is provided in the manuscript or supporting information (Supporting Information Table S4). Reported decomposition temperatures from sentences describing TGA traces were not used directly due to the ambiguities associated with authors reporting multiple temperatures for removal of guests and some authors reporting TGA decomposition temperature at complete collapse versus at onset. We instead extracted TGA traces from TGA-keyword-containing manuscripts, and we applied a consistent protocol for assignment of the TGA temperature from the TGA traces (Supporting Information Figure S4). Specifically, we quantified decomposition temperatures by identifying the temperature range corresponding to MOF decomposition, extracting tangent lines to the start and end of the decomposition process, and calculating their intersection point (Supporting Information Figure S4). For a fraction of manuscripts where sentences about TGA are present but no TGA trace could be identified, we did not assign a decomposition temperature (Supporting Information Table S7).

From the 3,809 downloaded manuscripts corresponding to at least one structure in the CoRE MOF database, 2,649 manuscripts contained keywords associated with solvent-removal stability. For this set, we identified 1,209 manuscripts corresponding to 2,290 structures that we could unambiguously label as stable or unstable. We then analyzed the data for duplicates using revised autocorrelation (RACs, see *Methods*) connectivity descriptors, although matching RAC descriptions could correspond to distinct MOFs that differ only by geometric properties (e.g., distinct pore volumes or crystal phases). If the identified MOF crystal structures had identical connectivity but conflicting stability labels, we removed all instances of that MOF from the dataset. We preserved instances of MOF connectivity duplicates with consistent stability labeling (Supporting Information Figure S5). After removing duplicate MOFs with conflicting labels, we obtained 2,179 structure–label pairs, a dataset we call the solvent-removal stability dataset (SSD). For the 3,809 downloaded manuscripts, 2,366 manuscripts with 4,366 structures contained TGA keywords within the main text of the manuscript, and 1,886 of these manuscripts had TGA traces with temperatures that we could extract. From this set of 1,886 manuscripts, we identified 3,132 thermal decomposition temperatures, a dataset we call the thermal stability dataset (TSD).

*MOF Solvent Removal Stability Classification*

For a range of applications from catalysis to gas adsorption, it is necessary to remove solvent and form active sites in porous MOFs[31] while preserving framework integrity in a process known as "activation". While some empirical rules for determining this stability after activation are known, e.g., that MOFs with long[64] and flexible[65] linkers or larger pores[66] are more likely to be unstable upon solvent removal, there are numerous exceptions. A recent study of MOFs with Zr secondary building units (SBUs) revealed that one large-pore MOF, PCN-57, is stable upon activation despite the fact that the smaller-pore MOF, UiO-67, collapses upon solvent evacuation[43]. No general rules have been established for how metal node, linker, and pore volume simultaneously influence a MOF's solvent-removal stability.

To develop a broader understanding of the principles that govern this aspect of MOF stability, we extracted literature data and assigned solvent-removal stability labels for the 2,179 MOFs in the SSD set (Supporting Information Figures S6–S8). Our curated data set has a slightly greater fraction (59%) of MOFs labeled as stable than unstable due to an expected bias for publishing successful results (Supporting Information Table S8). No single heuristic (e.g. pore volume or linker size) correlates strongly with solvent-removal stability over this set, motivating the development of machine learning (ML) classification models (Supporting Information Table S9).

Our goal is to build ML models that exhibit good performance for stability classification and maintain interpretability to provide insight into the key components of MOF chemistry and structure that determine solvent-removal stability. Thus, as in prior work[67-68] we employ feature selection in combination with kernel methods (e.g., Gaussian processes or GPs) because kernel methods inherently make predictions on new compounds informed by the similarity (i.e., distance in feature space) to known compounds. We also compare performance to alternative architectures (i.e., artificial neural networks, ANNs) that can yield better performance especially for large-scale chemical space exploration.[69] Our feature set in all cases consists of graph-based descriptors applied across the key domains of a MOF (i.e., SBU, linker, and functional groups) in addition to pore geometry descriptors (see *Methods*). The MOF-tailored[11], graph-based descriptors[67] were previously demonstrated to improve ML model performance for predicting properties and revealing design principles in gas adsorption.[11] Indeed, all models perform significantly better than random guess (AUC: 0.5) but less than a perfect classification (AUC: 1.0) on a set-aside test set. An ANN classifier performs comparably (AUC: 0.79) to a GP classifier trained on the most essential features (AUC: 0.81) and also affords greater interpretability (Figure 2 and Supporting Information Tables S10–S11 and Figure S9).

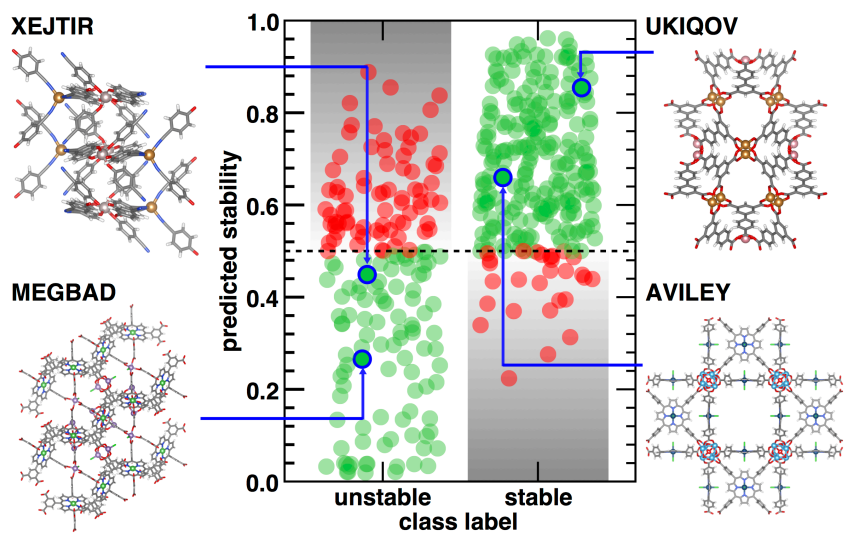

**Figure 2.** GP classifier model predictions vs actual class labels for MOFs in the test set for the solvent-removal stability data set. Data points are represented as translucent circles to depict data density and colored by the classification correctness: correct (green) and incorrect (red). Example structures and corresponding CSD refcodes for correct classifications are shown with blue outlines for two unstable MOFs: XEJTIR and MEGBAD, and two stable MOFs: UKIQOV and AVILEY. MOFs are shown in ball (metals) and stick representation, with Al in silver, Mn in purple, Co in pink, Ni in green, Cu in brown, Zn in pale blue, Hf in sky blue, Ir in teal, O in red, C in gray, N in blue, Cl in light green, and H in white.

Because the feature-selected GP classifies MOF solvent-removal stability reasonably well, we next analyzed the features that are most essential to these predictions. As could be expected, most (ca. 75%) of the selected features are related to linker chemistry (Figure 3 and Supporting Information Table S12). These linker features tend to emphasize the elemental identity and size of the substituents (i.e., chemical features) in the linker and the presence or placement of functional groups rather than the rigidity (i.e., topology and bonding) of the structures (Supporting Information Table S12). In comparison, only one geometric descriptor, the gravimetric surface area, is selected for predicting solvent-removal stability, likely in part because the selected graph descriptors that encode the linker and the number of atoms in the unit cell capture size effects indirectly (Supporting Information Table S12). In addition to these expected trends, the metal SBU chemistry contributes significantly (ca. 20% of features, Supporting Information Table S12). For the SBU features, elemental identity, relative size, and, to a lesser extent, electronegativity of the substituent metal node atoms are selected (Supporting Information Table S12). A simpler decision tree model also de-emphasizes pore size in classifying solvent-removal stability in an even more transparent fashion, despite not achieving the same performance as our GP classifier that can account for cooperativity among features for mapping structure–stability relationships (Supporting Information Figure S10). Overall, these observations highlight why simple heuristics for solvent-removal stability have thus far evaded the MOF community, as a range of chemical variations in the linker and metal node all play contributing roles in determining MOF stability upon activation.

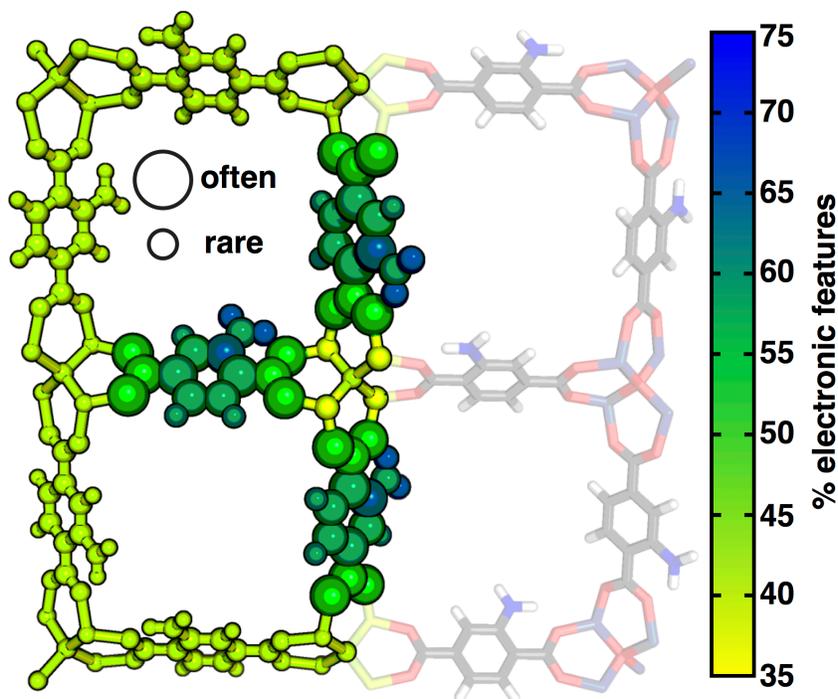

**Figure 3.** Schematic of relative importance and character of the selected feature set for predicting solvent-removal stability. Atom sizes of a representative MOF (here, MOF-5-NH$_2$) are scaled relative to the number of descriptor dimensions involving that atom. Features that describe the whole unit cell (full-scope unit cell RACs) are shown by the outer portion of the unit cell. Atoms are colored by the relative frequency of electronic features at each depth, with yellow indicating more geometric character, and blue indicating more electronic character. Atoms of MOF-5-NH$_2$ (right) are shown in a stick representation, with Zn in gray, O in red, C in gray, N in blue, and H in white.

To gain further insight into the structure–stability relationships encoded in our ML models, we analyzed representative examples of correct classification by the GP. In line with expectations that smaller, rigid ligands are stable upon solvent removal[67], the model confidently (i.e., based on classifier probability score) predicts a MOF with a symmetric rigid carboxylate linker (refcode: UKIQOV) to be stable (Figure 2). However, a model based only on a linker-rigidity heuristic would incorrectly classify another MOF (refcode: XEJTIR) with small linkers and mononuclear Cu/Al SBUs as stable. Instead, the ML model correctly predicts this MOF to be unstable (refcode: XEJTIR, Figure 2). Analyzing this MOF's nearest neighbors in the training data reveals that the ML model prediction is informed by mononuclear transition-metal SBUs that are unstable despite having rigid linkers (Supporting Information Figure S11). Thus, this analysis highlights that linker design for solvent-removal stability must be carried out with consideration of the MOF SBU chemistry (Figure 3).

By analyzing compounds with similar linker chemistry but distinct solvent-removal stability, we interrogated the metal dependence of our model's predictions. Overall, the key metal-centered (e.g., SBU) RAC features are more geometric than electronic in nature (Figure 3). Our classifier identifies a MOF with porphyrinic linkers and trinuclear Mn SBUs (refcode: MEGBAD) as unstable with respect to solvent removal because similar porphyrinic MOFs with trinuclear metal clusters also collapse upon activation (Figure 2 and Supporting Information Figure S11). For a MOF with porphyrinic linkers and a Hf-containing SBU (refcode: AVILEY), the model correctly predicts that this MOF will be stable with solvent removal (Figure 2 and Supporting Information Figure S11). The presence of some metals degrades the performance of the model. For example, we observe the least accurate predictions to be on 3$d$ transition-metal MOFs, likely due to oxidation- and spin-state dependent effects that can give rise to variations that are challenging for ML model predictions in comparison to heavier metals (Supporting Information Figure S12).

*MOF Thermal Stability Regression*

MOFs that demonstrate permanent porosity must be thermally stable for practical use. Thermogravimetric analysis (TGA) data is frequently utilized[38, 45] to quantify thermal degradation. Previous studies[34, 45] on MOFs have used TGA data to attempt to establish heuristics for their stability. The ready availability of TGA traces in most papers on MOFs make it possible for us collect a large number of thermal decomposition temperatures ($T_d$) and map them to MOF composition and connectivity. Over the 3,132 MOFs in the TSD with diverse chemistry (i.e., linkers, pore volumes, and metals), the $T_d$ values range from 35°C to 650°C, a far larger span than the typical uncertainty[45] (ca. 25°C) in a TGA measurement (Figure 4 and Supporting Information Figures S13–S15). We observe minor metal dependence in $T_d$, likely due to the presence of mixed-metal MOFs (Supporting Information Figures S13 and S16). As in the case of solvent-removal stability, we see no strong correlation between any single heuristic and $T_d$, again motivating ML models to predict thermal stability (Supporting Information Figure S16 and Table S13).

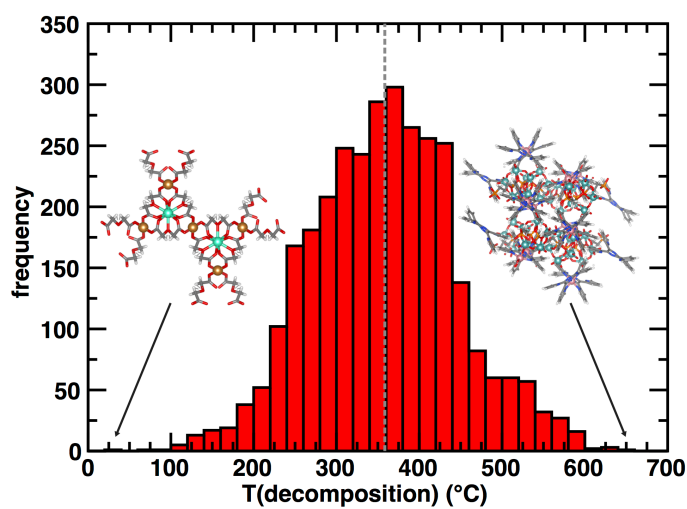

**Figure 4.** Histograms (bin width: 20°C) of thermal decomposition temperatures over the thermal stability dataset. The structures with the lowest (refcode: WEVQOD01) and highest (refcode: IFAREN) thermal stability are shown inset. MOFs are shown in ball (metals) and stick representation, with Co in pink, Cu in brown, Mo in blue-green, Gd in light green, O in red, C in gray, N in blue, P in orange, and H in white.

Both a feature-selected GP regression (GPR) model and an ANN exhibit good (mean absolute error or MAE: 44–47°C) performance on a set-aside test set (Figure 5 and Supporting Information Figure S16 and Table S14). Unlike in the case of solvent removal, prediction errors of $T_d$ are balanced across all metals (Supporting Information Figure S17). To put this performance in context, the errors are only somewhat larger than the typical uncertainty of the experimental measurement, while the ML model is both more rapid and accurate than possible with conventional simulation.

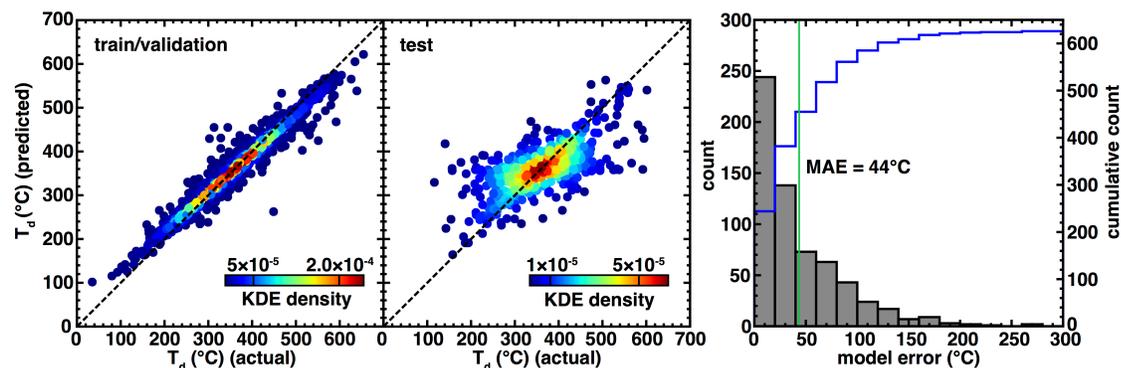

**Figure 5.** GPR model performance for predicting thermal decomposition temperatures in the thermal stability dataset. Parity plots (left) of actual vs predicted values for train/validation and test data points colored by kernel density estimation (KDE) density values, as indicated by inset color bars. In all cases, a black dashed parity line is shown. Distributions (right) of test set model errors for thermal decomposition temperatures (bins of 20°C), with the MAE annotated as a green vertical bar and cumulative count shown in blue according to the axis on the right.

Importantly, the selected features for the GPR again provide interpretation of what governs thermal stability. Analysis of these features indicates that both metal node and linker chemistry are important, with the latter selected with the highest frequency (Figure 6 and Supporting Information Table S15). Decomposing the selected features further, we observe that they are primarily electronic in character[67], i.e., most (ca. 65%) of them encode nuclear charge or electronegativity (Figure 6). This emphasis is even higher (ca. 73%) for the metal-centered feature subset (Figure 6 and Supporting Information Table S15). Unlike solvent removal, thermal stability is less dependent on the size of atoms around the metal (i.e., from *S* metal-centered RACs, see Computational Details). While both types of stability depend strongly on linker-focused features, the ones associated with functional groups that are especially distant from the metal matter most for thermal stability. These small differences do suggest some possibility for orthogonal design of the two types of stability, e.g., by altering electronic properties of the SBU for thermal stability and geometric properties for solvent removal stability (Figures 3 and 6).

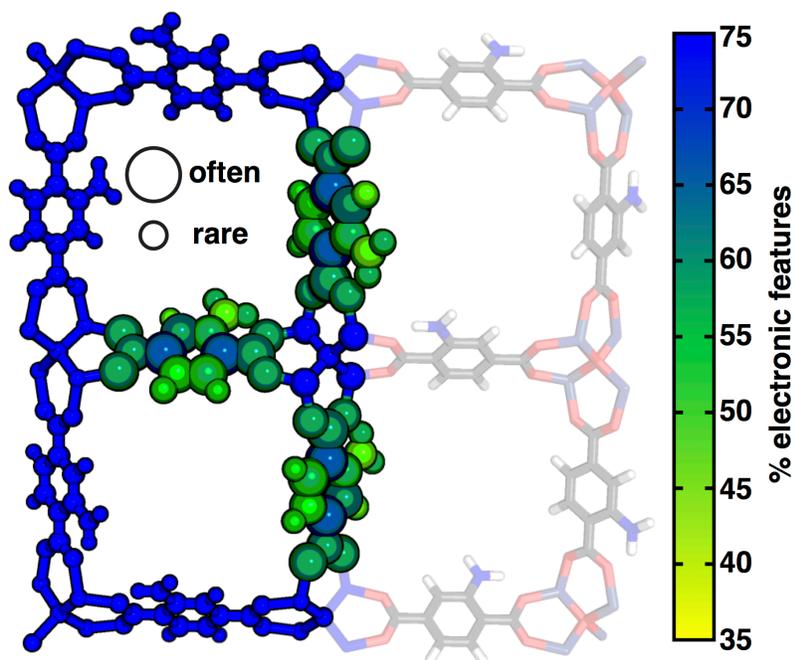

**Figure 6.** Schematic of relative importance and character of the selected feature set for predicting thermal stability. Atom sizes of a representative MOF (here, MOF-5-NH$_2$) are scaled relative to the number of descriptor dimensions involving that atom. Features that describe the whole unit cell (full-scope unit cell RACs) are shown by the outer portion of the unit cell. Atoms are colored by the relative frequency of electronic features at each depth, with yellow indicating more geometric character, and blue indicating more electronic character. Sets of features with > 75% electronic character are also colored in blue. Atoms of MOF-5-NH$_2$ (right) are shown in a stick representation, with Zn in pale blue, O in red, C in gray, N in blue, and H in white.

At a large bondwise distance from the metal in the SBUs, these and related geometric properties (i.e., identity and covalent radius) are present in selected features, indicating that linker rigidity influences thermal stability (Figure 6). Electronic contributions of metal chemistry instead likely alter the metal–ligand bond strength most, determining thermal stability. These statistically selected design principles support heuristics previously developed from such as hard-soft acid-base (HSAB) theory to rationalize the thermal stability of MOFs with high-valent metals[70] (e.g. Zr(IV)). As an example of linker chemistry importance, we observe significantly lower T$_d$ of a Zr MOF with piperazine phosphonic acid linkers relative to UiO-67, which uses biphenyl dicarboxylate linkers (Supporting Information Figure S18). While metal substitution can tune $T_d$, variation in linker chemistry is expected to exert a greater influence on thermal stability.

Specific thermal stability predictions by the GPR model illustrate these trends. The model accurately predicts a low $T_d$ (ca. 200°C) for a Cu MOF (refcode: RONSIY) with dibenzoate and bipyridine linkers, a prediction supported by the presence of a low-$T_d$ Co MOF in the training set (refcode: CUKVUA) containing similar linkers (Supporting Information Figure S19). Our model also accurately captures an increase in thermal stability (i.e., by > 100°C) with a more rigid benzene tricarboxylic acid linker (refcode: DOKHIV) in a Cu MOF, reinforcing the influential role of bonding within the linker on thermal stability (Supporting Information Figure S19).

Mn MOFs demonstrate a 400°C range for $T_d$ that can be captured by our model (Supporting Information Figures S16–S17). Within this fixed-metal subset, our model helps to identify how linkers impart thermal stability to exceptionally stable MOFs. The GPR model accurately predicts that a Mn MOF (refcode: JOYKUF) with biphenyl dicarboxylate linkers will be exceptionally stable ($T_d$: 427°C, Supporting Information Figure S19). The model also identifies that using a larger, bis(oxy)dibenzoate-substituted carboxylate linker (refcode: YINCIJ) for the same Mn SBU reduces $T_d$ by 100°C (Supporting Information Figure S19). Thus, the presence of oxygen atoms distal from the metal is identified by the GPR to reduce thermal stability. The GPR model also captures that exceptional thermal stability can be achieved by a MOF with a linker that is smaller and rigid (i.e., methylisophthalate, refcode: ZISFEO) instead, highlighting the model's emphasis of the number of bonds within the linker (Figure 6 and Supporting Information Figure S20). Employing rigid, aromatic linkers with fewer metal-distal heteroatom substitutions thus is a potential approach to promoting MOF thermal stability when holding the metal SBU fixed.

*Engineering Stability into Existing MOFs*

The approximately 5,800 synthesized MOF structures lacking automatically extracted stability measures provide a demonstration of the power of our predictive stability models. Applying the ML models only where they are reasonably confident[71-72], we are able to rapidly make a large set of blind predictions about this unseen data. In contrast, direct determination of each individual MOF's stability from this set would take a minimum of several minutes to hours for manual inspection of the manuscripts and supporting information to identify whether the quantities were present but automated extraction had failed. Alternatively, in cases where one or both quantities had not been determined, a new synthesis and measurement would be necessary, which could require days to weeks in the case of each MOF. Even with a conservative threshold on ML model domain of applicability, we predict both types of stability for nearly 1,500 of these MOFs for which our ML models are most confident in a total of an hour for the entire set, an acceleration that corresponds to several orders of magnitude over the fastest possible alternatives (Supporting Information Tables S16–S19 and Figure S23).

In addition, these predictions enable us to extend the abstract feature analysis to observations for a larger number of synthesized materials (Supporting Information Figures S21–S22). Over this set, these stability predictions highlight both intuitive and unexpected principles the models have learned. For example, our models predict the MOF with the largest pores in this experimental set is stable upon solvent removal, demonstrating the opportunities for this data-driven approach to identify and overcome limits of heuristics (Supporting Information Figures S24–S25). In terms of metal chemistry, MOFs containing alkali, alkaline earth, or heavy (e.g. period 5 or 6) metals are most likely to exhibit both measures of stability (i.e., thermal and solvent removal), with lighter Zn MOFs as the primary exception to this rule (Figure 7 and Supporting Information Table S19 and Figure S26). Nearly all 3*d* transition metals and late 4*d* transition metal Cd are the most frequently occurring metals in MOFs predicted to have good solvent stability but poor thermal stability. Because metal identity is not the sole decider of both stability measures, numerous solvent-removal stable but thermally unstable exceptions, e.g., with alkali and alkaline earth metal MOFs, are also predicted (Supporting Information Figure S26).

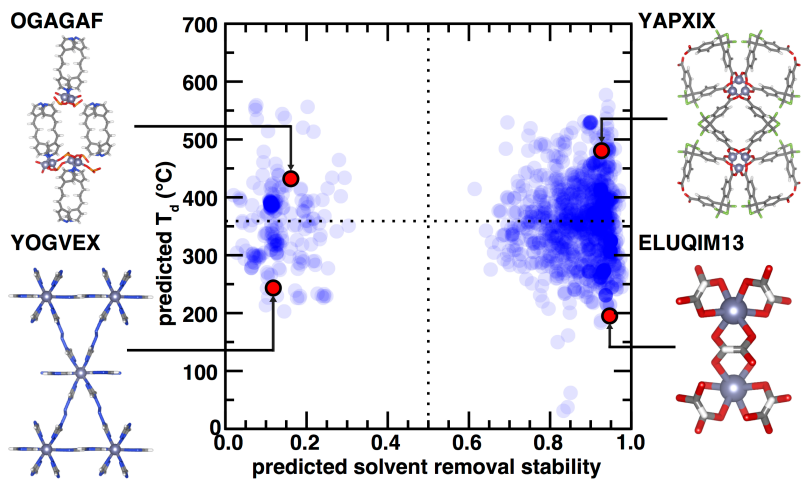

**Figure 7**. Predicted solvent-removal stability (probability, no units from 0 for unstable to 1 for stable) and thermal stability ($T_d$, in °C) for the featurizable CoRE MOF data set for MOFs that fall within UQ cutoffs for both models. Points are shown as blue translucent circles. Four representative Zn MOFs are shown as red circles, with phosphonate linkers (refcode: OGAGAF, top left), diazene linkers (refcode: YOGVEX, bottom left), carboxylate linkers (refcode: YAPXIX, top right), and formate linkers (refcode: ELUQIM13, bottom right). MOFs are shown in ball (metals) and stick representation, with Zn in gray, O in red, C in gray, N in blue, F in light green, P in orange, and H in white.

We expect linker chemistry to play a critical role in MOF stability, but the specific role of the linker is challenging to uncover. Instability could be due to weaker intralinker covalent bonds or to weak non-covalent interactions that limit the greater structural stability. Indeed, specific elements (i.e., Cl, B, or Si) are often present in MOFs that are thermally unstable and occasionally also unstable with solvent removal, whereas all fluorine-containing MOFs are more stable (Supporting Information Figures S27–S29). Oxygen coordinating-atom (e.g. carboxylate) linkers form more stable MOFs than nitrogen coordinating-atom (e.g. tetrazole or pyridyl) counterparts (Supporting Information Figures S30–S31). Although these rules generally hold, several phosphonate metal-coordinating MOFs are predicted by both ML models to be stable, despite expectations of reduced thermal stability[45] (Supporting Information Figure S32). While naphthalene dicarboxylate linkers make stable MOFs, our models distinguish a MOF with an isomer of naphthalene dicarboxylate (refcode: HEBTAL) and predict it be unstable upon solvent removal (Supporting Information Figures S30 and S33). Because we could not automatically download and mine this article due to the publisher's restrictions, our prediction of the stability of this MOF represents a naturally blinded test for our model. Subsequent manual inspection of the literature[73] confirmed that our ML model correctly predicted this MOF's collapse upon solvent removal (see *Methods*).

Feature analysis had indicated opportunities for orthogonal design of stability measures (e.g., the higher importance of distant functional groups for thermal stability vs metal-proximal chemistry for solvent removal), suggesting that ML models can be exploited to rapidly identify the necessary changes to increase one metric without worsening the other. A representative Zn MOF with dimethoxy-functionalized biphenyl dicarboxylate linkers (refcode: XAMHEA) is thermally stable but is unstable with solvent removal (Figure 8). Feature selection had suggested that metal-distal linker heteroatoms could be a source of reduced solvent removal stability, motivating elimination of the methoxy groups on the linker while maintaining the underlying MOF connectivity. Indeed, this changes the predicted solvent-removal stability from unstable to stable (0.11 vs 0.89) while preserving the prediction of good thermal stability (433°C vs 430°C, Figure 8). Replacing all Zn atoms with Ca (i.e., from hard to intermediate softness) that might be expected to influence stability has no effect, but removing the methoxy groups and substituting with Ca at the same time indeed improves both the predicted solvent removal and thermal stabilities (Figure 8).

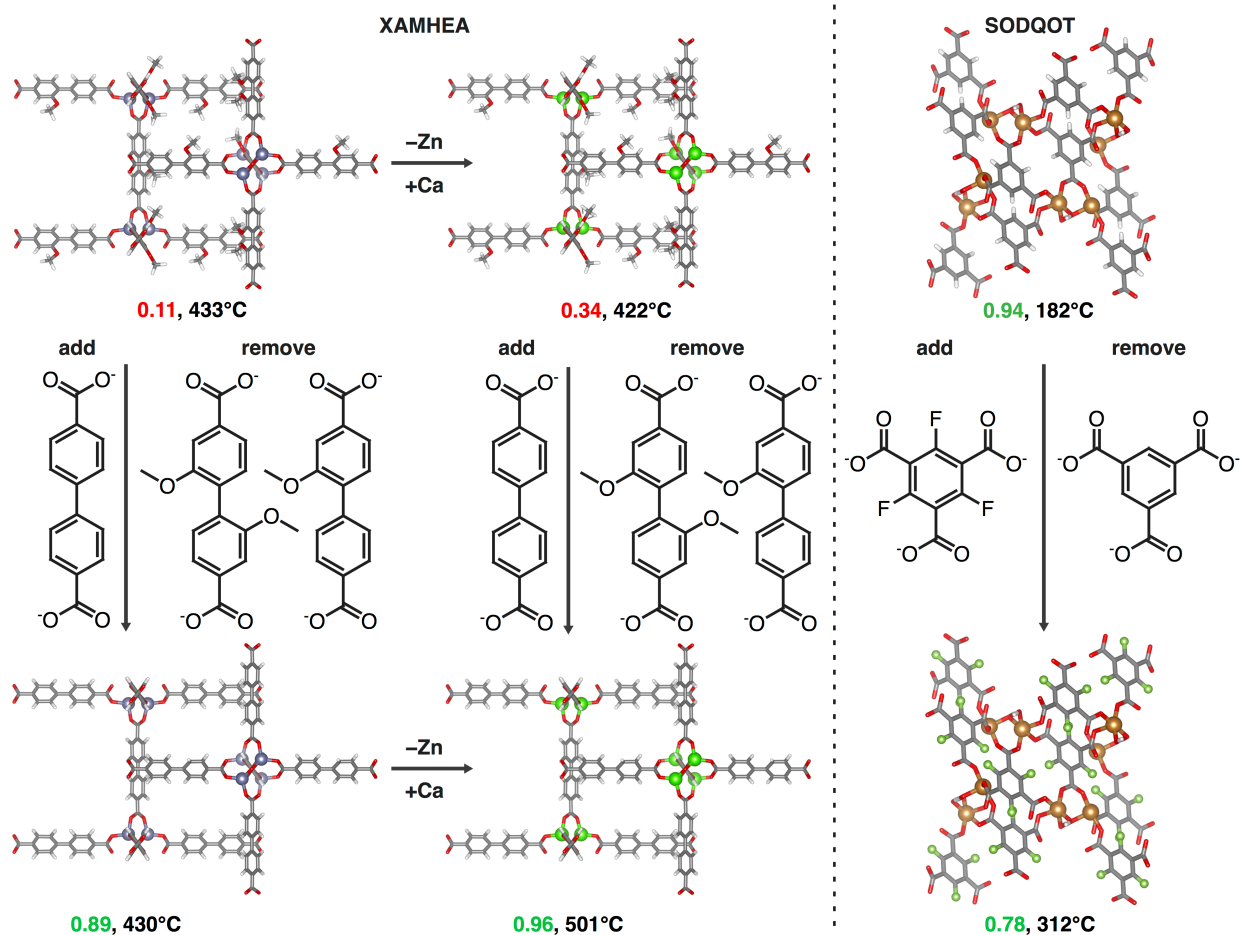

**Figure 8**. Schematic of structural changes that are predicted to enable independent tuning of solvent-removal stability and thermal stability for two MOFs. For each MOF, the predicted solvent-removal stability (from 0 for unstable in red to 1 for stable in green) and predicted $T_d$ (in °C) are shown. Alterations of the starting Zn MOF (refcode: XAMHEA, top left) include replacing Zn with Ca (top middle), removing methoxy groups from linkers (bottom left), or both (bottom middle). Also shown are changes to the linker of a Cu MOF (refcode: SODQOT, top right) by fluorination of the linker (bottom right). MOFs are shown in ball (metals) and stick representation, with Zn in light blue, Ca in bright green, Cu in brown, O in red, C in gray, F in green, and H in white.

Because 3$d$ transition-metal MOFs are useful for catalysis, we demonstrate how these models and the principles that they reveal can be used to engineer thermal stability into a MOF that is already stable upon solvent removal. We start with a representative Cu MOF with benzene tricarboxylic acid linkers (refcode: SODQOT) that lacks thermal stability and would degrade under catalytic conditions for light alkane oxidation[28, 74] ($T_d$: 182°C, Figure 8). From analyses of linker elemental distributions, fluorinating the linker is expected to improve thermal stability (Supporting Information Figure S27). Indeed, we find that our model predicts we can improve $T_d$ by 130°C without compromising solvent removal stability, which would make this MOF suitable for catalysis.[28, 74] The improvement in thermal stability is likely due to the more electron deficient fluorinated linker enabling stronger metal–ligand bonds (Figure 8 and Supporting Information Figure S34). Our data-driven insights provide a path forward to improve MOF stability, thus broadening their use as catalysts and other applications that require robust materials performance under harsh conditions.

*Conclusions*

Although the tailored metal active sites and porous architectures of MOFs hold great promise for engineering challenges ranging from gas separations to catalysis, a lack of understanding of how to improve their stability limits their use in practice. From the available literature, we have extracted thousands of reports of the key aspects of MOF stability necessary for their practical application: the ability to withstand high temperatures without degrading and the capacity to be activated by removal of solvent molecules without collapsing. Analysis of the relationships between stability properties and the associated MOF chemical and geometric structures in this set reveals that existing heuristic models derived from smaller sets of MOFs within the same family will fail when applied more globally. We trained predictive ML models to encode the structure–property relationships for both thermal and solvent-removal stability measures using graph- and pore-structure-based representations. These trained models predict both quantities orders of magnitude faster than conventional physics-based modeling or experiment.

Interpretation of ML model predictions highlights the distinct roles of linker connectivity (i.e., the degree of bond saturation) in determining solvent-removal stability and linker chemistry in determining thermal stability. The models are also useful in testing heuristics and assumptions about MOF stability. While intuition suggests MOFs with large pore volumes should be unstable, we confirmed model predictions of stability for specific large pore volume MOFs against literature reports. Expanding our analysis to all MOFs in the CoRE MOF database for which automated extraction of both thermal and solvent stability information was not available highlighted important trends, such as that linkers that coordinate the metal with oxygen atoms are typically much more stable than nitrogen-coordinating linkers that might normally be preferred for catalytic applications.

Finally, we leveraged these predictive models to identify strategies to engineer MOF structures to increase their stability. For the $3d$ transition-metal-containing MOFs that are target for catalysis but frequently unstable upon heating, we identified linker chemistry modifications that could increase the thermal stability necessary for catalytic applications. These first-generation ML models were naturally constrained in their performance, in part by the representations or ML models themselves but primarily by the amount of data that could be easily extracted from the literature. Toward that end, we have provided our ML models and extracted properties from the data sets in a user-friendly web interface, which welcomes both feedback and the incorporation of additional community data[75]. We expect our approach, including natural extensions to other properties, to accelerate the time to discovery of stable, practical MOF materials by both computational and experimental researchers.

*Methods*

*Software and Workflows*

We used the Cambridge Structural Database (CSD) v5.41 python application programming interface (API)[76] to obtain article digital object identifiers (DOIs) for each MOF as identified by a unique CSD refcode. We used a locally modified version of the ArticleDownloader toolkit[77] to automate article downloads and construct a corpus of MOF articles from various publishers. We performed direct downloads of articles published by the Royal Society of Chemistry (RSC), Wiley, the American Association for the Advancement of Science (AAAS), Springer, and Nature Publishing Group, while obtaining articles published by the American Chemical Society (ACS) via a direct download agreement. We parsed all documents with the ChemDataExtractor toolkit[78] to perform tree-based searches on structured HTML and XML documents. Using the Stanza Natural Language Processing (NLP) toolkit[63], we performed dependency parsing on sentences tokenized by ChemDataExtractor, to disambiguate sentences with complex sentence structure. We digitized TGA traces from main text figures and supporting information using the Webplotdigitizer tool[79], to identify thermal decomposition onset temperatures. Detailed results of digitized TGA traces are provided in the Supporting Information.

*Machine Learning Models and Representations*

Graph-based revised autocorrelations (RACs)[67, 80-81] for each MOF were computed using molSimplify[11, 82]. Geometric properties (e.g. pore volume and gravimetric surface area) were computed using the Zeo++ code, using a nitrogen probe molecule[83-84]. We used RAC features supplemented with geometric features as inputs to all models, as done in our prior studies on MOFs[11]. RACs are sums of products and differences of five heuristic atomic properties: nuclear charge ($Z$), identity ($I$), topology ($T$), electronegativity ($\chi$), and covalent radius ($S$), taken up to a maximum bond path ($d$) of three bonds. From a set of 160 possible RACs, 134 remain after the elimination of invariant RACs, which were then supplemented with 14 geometric features computed by Zeo++, which quantify the pore size, surface area, and void space of each MOF (Supporting Information Tables S20–S21). We scaled all data to zero mean and variance of one prior to training models.

We trained all machine learning (ML) models using GPy[85], scikit-learn[86], and keras[87] with a tensorflow[88] backend to independently classify the stability with respect to solvent removal and predict the thermal decomposition temperature. We performed hyperparameter optimization for kernel-based models using exhaustive grid search, and artificial neural networks (ANNs) using Hyperopt[89], with a random 80%/20% train/test split, with 20% of the training data (16% overall) set aside as a validation set used for hyperparameter selection (Supporting Information Table S22). As in prior studies[67-68], we carried out feature selection on kernel ridge regression (KRR) models, and repeated this process for Gaussian process regression (GPR), support vector classification (SVC), and Gaussian process classification (GPC) models, using random-forest[90]-ranked recursive feature addition[68] (RF-RFA). For KRR, GPR, SVC, and GPC models, we trained classification and regression models with radial basis function (RBF) kernels. In addition, for the GPR and GPC models, we trained classification models with Matern-3/2 and Matern-5/2 kernels. With each feature addition, we evaluated the error metric on the validation set (e.g. mean absolute error (MAE) for regression tasks, and area under the receiver operating characteristic curve (AUC) for classification tasks) after retraining the models with the new feature. We stopped RF-RFA after the validation metrics stopped improving. We trained final models on the full training set (e.g. 80%), and evaluated their performance on the set-aside test set.

ASSOCIATED CONTENT

**Supporting Information**. The following files are available free of charge.

Data filtering statistics for CoRE MOF featurization; publisher dependent statistics on manuscripts; keywords used to quantify stability; relative location distribution for solvent removal stability sentences; examples of dependency parsing and keyword matching; keywords used for eliminating other types of stability; structure-paper mapping counts; example of thermogravimetric analysis and corresponding statistics for gathered data; features used for ML model development and corresponding hyperparameters; pore size distributions, metal distributions, and linker distributions of solvent removal stability and thermal stability data sets; distribution of text mined flag labels; correlation between continuous RAC descriptors and binary stability flag; ML model performance for all models before and after feature selection; binary tree classification using linker and geometric features; prediction accuracy for classification by metal; selected features for solvent removal and thermal stability; correlation between RAC features and thermal decomposition temperature; examples of Zr MOFs with differing linkers; analysis of nearest neighbor compounds of representative compounds of the thermal stability and solvent removal stability data sets; distribution of CoRE MOF predictions for compounds without a ground truth; principal component analysis comparing the design space with no ground truth to the underlying model training data for solvent removal and thermal stability; latent space uncertainty quantification for both classification and regression; latent space entropy and distance cutoffs; description of ranges of solvent removal and thermal stability and quantification of metal diversity in each region; profile of time necessary to make prediction; example stable phosphonate MOFs predicted by models; principal component analysis of geometric space; linker composition by stability classification; most prevalent atoms by stability classification; principal component analysis of linker chemistry and comparison to expectations from heuristics; frequency of linkers by stability classification; histograms of maximum included sphere by stability classification; quantification of thermal stability and solvent removal stability colored by electronegativity differences (PDF)

Extracted labels and sentences for the solvent removal stability and thermal stability data sets; RAC and geometric descriptors and train, validation, and test set splits for all compounds in the solvent removal stability and thermal stability data sets; identification of duplicate compounds in the solvent removal data set; model weights for solvent removal stability and thermal stability ANNs (ZIP)

AUTHOR INFORMATION

**Notes**

The authors declare no competing financial interests.

ACKNOWLEDGMENT

The authors acknowledge primary support by DARPA (grant number DE18AP00039) for the text extraction and machine learning efforts. MOF analysis and property selection was supported by the U.S. Department of Energy under Grant Number DE-SC0012702 (to A.N.). Some of the database curation efforts were supported by the Office of Naval Research under grant number N00014-20-1-2150 (to A.N., C.D. and H.J.K). This work was also partially supported by a National Science Foundation Graduate Research Fellowship under Grant #1122374 (to A.N.). H.J.K. holds a Career Award at the Scientific Interface from the Burroughs Wellcome Fund, an AAAS Marion Milligan Mason Award, and an Alfred P. Sloan Fellowship in Chemistry, which supported this work. The authors thank Adam H. Steeves and Vyshnavi Vennelakanti for providing a critical reading of the manuscript. The authors also acknowledge helpful conversations with members of the Elsa Olivetti lab.

**TOC GRAPHICS**

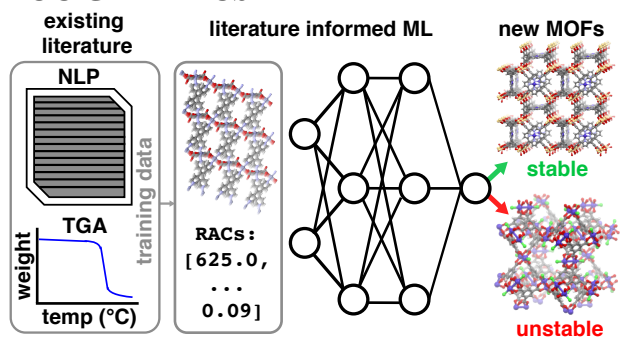

# Supporting Information for
## *Using Machine Learning and Data Mining to Leverage Community Knowledge for the Engineering of Stable Metal-Organic Frameworks*

Aditya Nandy[1,2], Chenru Duan[1,2], and Heather J. Kulik[1]
[1]Department of Chemical Engineering, Massachusetts Institute of Technology, Cambridge, MA 02139
[2]Department of Chemistry, Massachusetts Institute of Technology, Cambridge, MA 02139

**Contents**

**Table S1**. Statistics on structure attrition for MOF RAC featurization, including reasons for featurization failure. The starting set of structures is the 10,143 structures in the all solvent removed (ASR) subset of the CoRE MOF database. MOF RACs could be generated from cif files of 9,597 MOFs. Failure modes include: residual floating solvent indicating that the sanitized structure remains unclean; overlap between atoms, indicating poor resolution of the crystal structure; irreducible unit cell, indicating large unit cell lacking symmetry; no metal, indicating the lack of a metal in a cif file; tiny linker, indicating unclear distinction between metal-local and linker chemistry; and geometric property failure, indicating a failure of Zeo++ to obtain geometric properties of the MOF.

| starting structures | floating solvent | atomic overlap | irreducible unit cell | no metal | bulk metal / tiny linker | geometric property failure | final |
|---|---|---|---|---|---|---|---|
| 10,143 | 453 | 20 | 35 | 10 | 25 | 3 | 9,597 |

**Table S2**. Statistics on the number of papers that could be downloaded for a given publisher, along with the DOI prefix for that publisher and the counts in both the full dataset and the subset of 1:1 paper:structure maps. The majority of downloadable articles are derived from the American Chemical Society (ACS), the Royal Society of Chemistry (RSC), and Wiley, with fewer coming from Nature publishing group, Springer, or the American Association for the Advancement of Science (AAAS). Springer has three distinct DOI prefixes for different journal subsets. The 3,809 manuscripts in the full set represent 7,004 distinct MOF structures as denoted by their CSD refcode.

| publisher | prefix | count in full DOI set | number of featurizable structures |
|---|---|---|---|
| RSC | 10.1039 | 1,560 | 2,658 |
| ACS | 10.1021 | 1,438 | 2,770 |
| Wiley | 10.1002 | 638 | 1,211 |
| Springer | 10.1007 | 91 | 113 |
| Nature | 10.1038 | 50 | 173 |
| AAAS | 10.1126 | 18 | 59 |
| Springer | 10.1134 | 9 | 13 |
| Springer | 10.1023 | 5 | 7 |
| Total | - | 3,809 | 7,004 |

**Table S3**. Statistics over the full MOF set for structures that could not be automatically downloaded for text mining. The majority of non-downloaded articles are from Elsevier, International Union of Crystallography (IUCr), Taylor and Francis, or Chemical Society of Japan journals, with smaller journals appearing less frequently. Each set of manuscripts corresponds to a distinct number of crystal structures for which descriptors can be obtained.

| publisher | prefix | count in full DOI set | number of featurizable structures |
|---|---|---|---|
| Elsevier | 10.1016 | 888 | 1,250 |
| IUCr | 10.1107 | 154 | 173 |
| Taylor and Francis | 10.1080 | 101 | 123 |
| Elsevier | 10.1006 | 36 | 37 |
| Chemical Society of Japan | 10.1246 | 34 | 42 |
| De Gruyter | 10.1524 | 18 | 18 |
| Chinese Journal of Structural Chemistry | 10.14102 | 15 | 17 |
| Wiley | 10.1002 | 14 | 20 |
| CSIRO | 10.1071 | 13 | 24 |
| Chinese Journal of Inorganic Chemistry | 10.11862 | 11 | 15 |
| Bulletin of the Korean Chemical Society | 10.5012 | 9 | 18 |
| Acta Chemica Scandinavica | 10.3891 | 7 | 7 |
| De Gruyter | 10.1515 | 6 | 6 |
| MDPI | 10.3390 | 4 | 4 |
| Proceedings of the National Academy of Sciences | 10.1073 | 3 | 7 |
| Chemical Journal of Chinese Universities | 10.7503 | 2 | 2 |
| SIOC Journals | 10.6023 | 2 | 3 |
| Scientific Research Publishing | 10.4236 | 2 | 3 |
| SAGE Journals | 10.3184 | 2 | 2 |
| Trans Tech Publications | 10.4028 | 2 | 2 |
| Russian Journal of Chemistry | 10.7868 | 2 | 4 |
| Cambridge University Press | 10.1017 | 2 | 5 |
| Other publishers | 10.3969, 10.21060, 10.1135, 10.6060, 10.1179, 10.16533, 10.1142, 10.2478, 10.5560, 10.2298, 10.3866, 10.1154, 10.14233, 10.2116, 10.1360, 10.1139 | 1 each 16 total | 23 |
| Total | - | 1,343 | 1,805 |

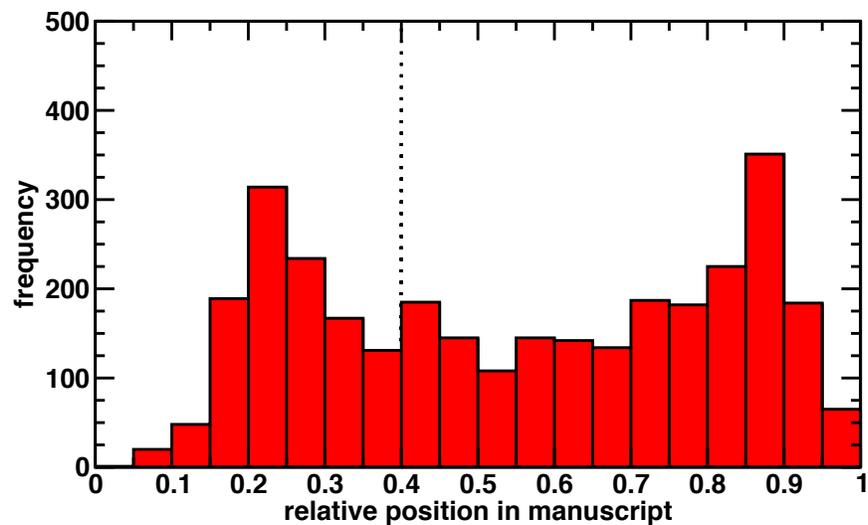

**Figure S1**. The relative position of all sentences identified as relevant to solvent-removal stability for manuscripts where an introduction cannot be explicitly identified. The relative position is defined as the sentence number within the document, normalized by the number of sentences in the document. Based on the distribution of positions in which solvent-removal stability sentences appear, we select a cutoff of 0.4 to eliminate sentences that are expected to represent introductory material. The histogram spans the range 0 to 1, with bin widths of 0.05.

**Table S4**. Keywords used to identify a MOF for solvent-removal stability. Stemmed word forms were compared to obtain fuzzy matches (via substring matching) to the text. For solvent-removal stability, one or more solvent keywords must be matched in addition to one or more stability or collapse keywords. Stability and collapse keywords were checked independently due to dependency parsing and the impact of negation on the assigned label of the sentence. Keywords used to identify manuscripts with TGA traces are also denoted.

| collapse keyword | stemmed form for search |
|---|---|
| collapse / collapsing / collapsed | collaps |
| deform / deforming / deformed / deformation | deform |
| amorphous / amorphize | amorph |
| blockage | blockage |
| degrade / degrading / degraded / degradation | degrad |
| unstable | unstable |
| instability | instability |
| destroy | destroy |
| one step weight loss / one-step weight loss | one step weight |
| single step weight loss | single step weight |
| **stable keyword** | **stemmed form for search** |
| stable | stable |
| stability | stability |
| preserve / preserving / preserved | preserv |
| crystallinity | crystallinity |
| coordinatively unsaturated / coordinatively unsaturating | coordinatively unsaturat |
| porosity | porosity |
| microporosity | microporosity |
| retain / retaining / retained | retain |
| maintain / maintaining / maintained | maintain |
| two step weight loss / two-step weight loss | two step weight |
| **solvent keyword** | **stemmed form for search** |
| solvent / solvents | solvent |
| guest / guests | guest |
| desolvate / desolvating / desolvated | desolv |
| remove / removes / removing / removed | remov |
| capillary | capillary |
| activate / activating / activation / activated / activates | activat |
| evacuate / evacuating / evacuation / evacuated / evacuates | evacuat |
| dehydrate / dehydrating / dehydration / dehydrated / dehydrates | dehydrat |
| eliminate / eliminating / elimination / eliminated / elminates | elminat |
| water | water / H2O |
| DMF | DMF / formamide |
| DMA | DMA / methylamine / diamine |
| EtOH / MeOH | EtOH / MeOH / ethanol / methanol |
| pyrrolidone | pyrrolidone |
| **TGA keyword** | **stemmed form used for search** |
| thermogravimetric analysis | thermogravimetric analysis / TGA |
| thermal-gravimetric analysis | thermal-gravimetric analysis |
| thermal gravimetric | thermal gravimetric / TG |

**Table S5**. Examples of keyword matches obtained from manuscripts. All manuscripts are checked for solvent and collapse/stability keywords. These sentences are then passed to syntactic dependency parsing and assigned a label. The -1 label corresponds to a negative mention and +1 for a positive mention. These scores are used to determine if the overall labeling of the material should be positive or negative.

| sentence (doi) | solvent keyword | stability / crystallinity keyword | assigned label |
|---|---|---|---|
| The crystal structure of this material shows very large void spaces, but solvent-exchange and evacuation led to the collapse of the structure and a loss of crystallinity. (10.1039/C5CE00886G) | evacuation | collapse | -1 |
| Removal of this solvent from within the pores may be effected thermally and leads to framework collapse. (10.1039/b009455m) | solvent | collapse | -1 |
| This comparison showed that the framework collapses after evaporation of the solvent guest molecules, which is not surprising since the coordination polymer is only two-dimensional (see above). (10.1002/ejic.200800222) | solvent / guest molecules | collapse | -1 |
| The crystals of the present coordination polymers were found to lose their crystallinity when exposed to air presumably due to loss of water of crystallization and decompose in the temperature range of 208-250°C. (10.1021/ic0204777) | water | crystallinity (negated) | -1 |
| The peaks of the experimental diffraction pattern of 2 are broad compared to the predicted pattern, and a loss of crystallinity is credited to structural collapse upon desolvation of the bulk material.(10.1021/ic401305c) | desolvation | collapse | -1 |
| Samples of 1 were heated at 180 °C for 12 h and the dehydrated samples showed no obvious changes in the PXRD pattern, which suggested no structural transformation upon removal of guest water molecules. (10.1039/C3CE41008K) | dehydrated / guest molecules | structural transformation (negated) | 1 |
| Upon desolvation, the elongated, partially flexible ligand can distort to lower the void space in the crystal and enhance stability. (10.1021/ic202429) | desolvation | stability | 1 |
| Powder X-ray diffraction (PXRD) data confirmed the purity of the bulk sample of 1 and the stability of the activated one (Fig. S5). (10.1039/c2ce06384k) | activated | stability | 1 |
| Moreover, from these studies, compound 1 is one of the few compounds that maintains its structural integrity after the removal of guest molecules. (10.1002/1099-0682(200203)2002:4<797::AID-EJIC797>3.0.CO;2-V) | guest molecules | integrity | 1 |
| The PXRD patterns of the activated samples (details of the procedure for activation of the MOFs are given in the Experimental Section) indicated stability of their networks even after removal of the solvent molecules from the pores. (10.1002/chem.201302455) | activated / solvent molecules | stability | 1 |

This prevents an irreversible pore system collapse that could otherwise occur during harsh solvent evacuation.

| dependency pairs | |
|---|---|
| (prevents, This) | (occur, could) |
| (ROOT, prevents) | (occur, otherwise) |
| (collapse, an) | (collapse, occur) |
| (collapse, irreversible) | (evacuation, during) |
| (collapse, pore) | (evacuation, harsh) |
| (collapse, system) | (evacuation, solvent) |
| (prevents, collapse) | (occur, evacuation) |
| (occur, that) | (prevents, .) |

Unstable keyword (collapse) identified in addition to solvent keyword (evacuation), and negation (prevents) identified, allowing correct label assignment.

**Figure S2.** Example of dependency parsing on a sentence identified in DOI 10.1021/acs.inorgchem.6b00530 that would be misidentified by exact phrase matching. Regular expression-based searches identify the sentence as pertaining to solvent evacuation stability, due to the presence of collapse keywords and solvent keywords. This sentence is further broken down into dependency pairs, and all dependencies are analyzed for negations (e.g. prevents collapse). Identification of the negations enables assignment of the correct solvent evacuation stability label. The collapse keyword is shown in gray, solvent keyword in blue, and negation in red. Dependency parsing also identifies the root of the sentence (e.g. the main action). For simple negations (e.g. "not stable"), dependency parsing would form a pair (not, stable) that will be consistently identified, even in the presence of modifier words that would cause exact matching to fail (e.g. "not very stable").

**Table S6**. Examples of keyword matches obtained from manuscripts that are false positives regarding water stability or air stability. All matches for stability are checked for these phrases to ensure that the sentence is not about these other classes of stability.

| type of stability | common phrases to identify false positives |
|---|---|
| air | air stability, air-stability, air stable, air-stable, exposure to air, stable to air, stability to air |
| water | water stability, water-stability, water stable, water-stable, exposure to water, stable to water, stable in water, water unstable, water instability, stability to water, hydrothermal stability |

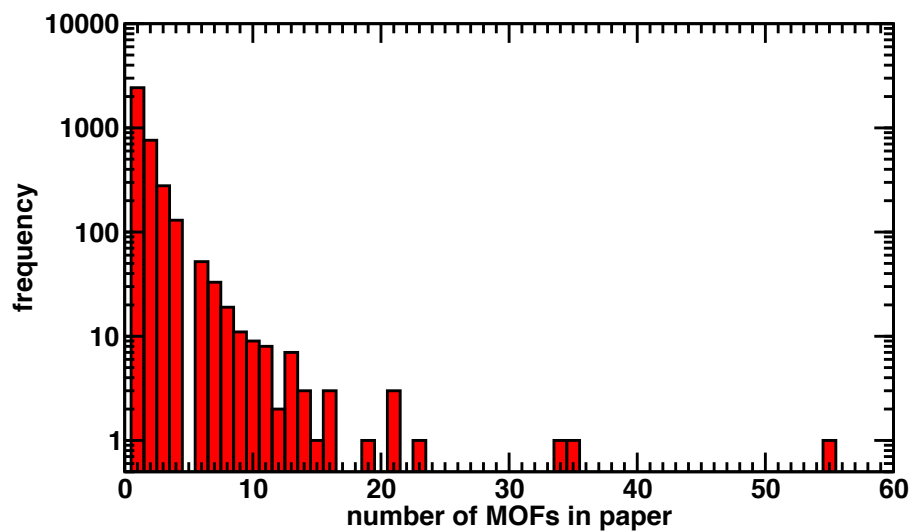

**Figure S3.** Bar plot of the frequency of finding a number of structures from the CoRE MOF dataset within a single manuscript. Although certain manuscripts may have multiple deposited structures, only a subset of those structures may be in the CoRE MOF dataset. The number of MOF structures within each plot is denoted by each bar, with a minimum of one structure per paper, and a maximum of 55 structures in one paper. The y-axis is shown on a log scale to highlight the distribution of MOFs per paper.

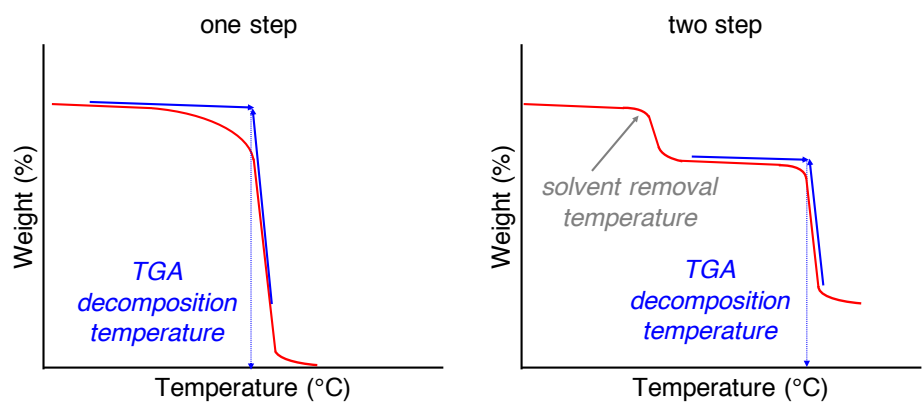

**Figure S4.** Hypothetical thermogravimetric analysis (TGA) traces of MOFs showing thermal decomposition temperatures. During a TGA experiment, the temperature is increased, and the mass and identity of compounds removed from the sample are analyzed. Compounds that are not stable with respect to solvent evacuation will demonstrate a single-step TGA where the MOF decomposes upon solvent removal (left). MOFs that are stable with respect to solvent removal will show a distinct temperature for solvent removal, and a plateau of stability before the onset of decomposition (right). Decomposition temperatures are measured from TGA traces by identifying the intersection of tangent lines, and reading the corresponding temperature, as denoted inset by a dotted blue arrow.

**Table S7.** The number of compounds for which a TGA keyword exists within the manuscript (excluding introductions), compared to the number of cases in which a TGA trace can be identified within the manuscript, followed by the number of compounds that can be mapped to a TGA decomposition temperature. The initial set of manuscripts is the 3,809 manuscripts that are tied to 7,004 featurizable MOF structures. The final curated set consisted of 3,132 thermal decomposition temperatures from 1,886 manuscripts, as indicated in the bottom row of this table.

| elimination step | number of manuscripts remaining (eliminated) | number of MOFs remaining (eliminated) |
|---|---|---|
| TGA keyword within manuscript | 2,366 (1,443) | 4,366 (2,638) |
| TGA figure isolated and could be mapped back to structure from CoRE MOF database | 1,886 (480)<br><br>breakdown of 1,886 manuscripts:<br>309 with a TGA plot in main<br>1,590 with a TGA plot in SI<br>13 with both (for different structures) | 3,132 (1,233)<br><br>breakdown of 3,132 structures:<br>410 reported in main<br>2,722 reported in SI |

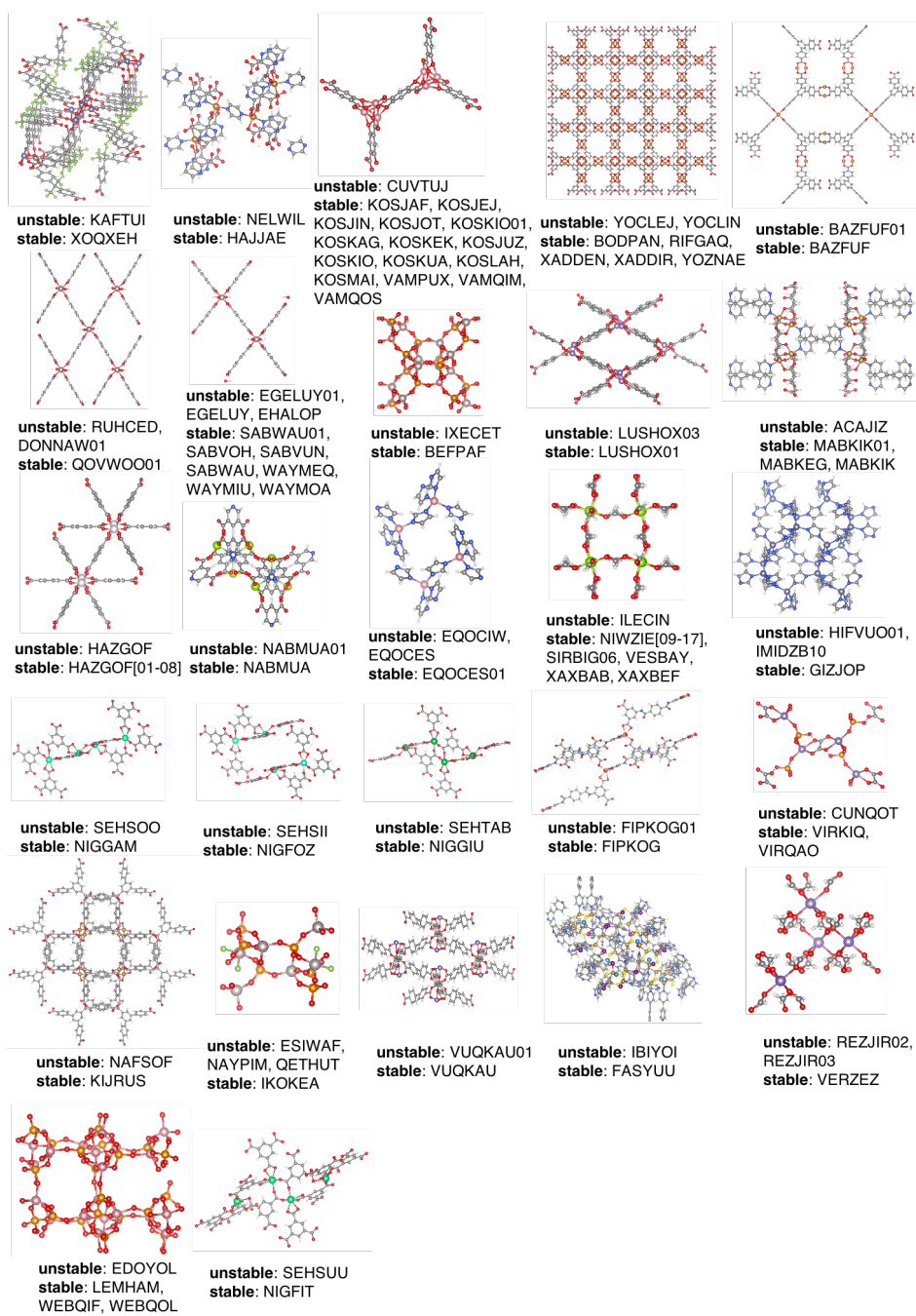

**Figure S5.** The 27 MOFs that contain duplicate entries (111 MOF entries total) with conflicting labels for solvent-removal stability and that were removed from the SSD data set. CSD refcodes corresponding to MOFs labeled as stable and unstable respectively are denoted under each set of duplicates. Unit cells are colored as follows: gray for carbon, white for hydrogen, red for oxygen, blue for nitrogen, orange for phosphorus, yellow for sulfur, lime for fluorine, dark blue for zinc, brown for copper, pink for cobalt, dark pink for gallium, silver for aluminum, yellow-green for magnesium, purple for manganese, light silver for scandium, dark teal for tungsten, light green for holmium, light teal for terbium, dark green for ytterbium, dark teal for erbium.

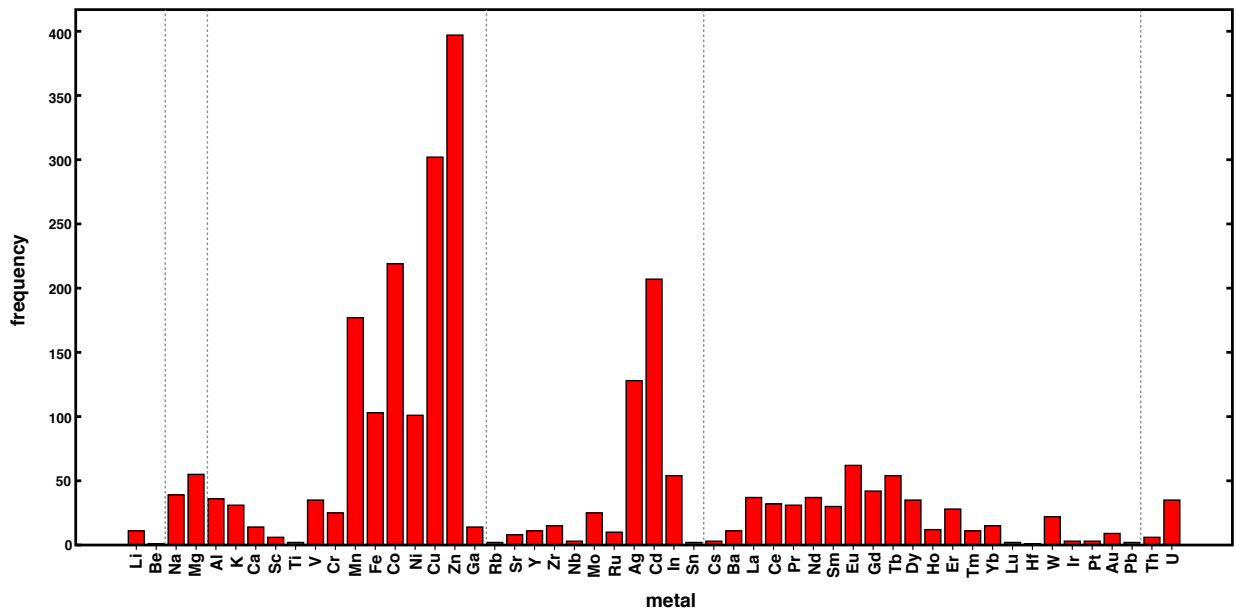

**Figure S6**. Frequencies of the metals that appear in the 2,179 MOFs in the solvent-removal stability data set. The majority of MOFs (83%, 1,815 out of 2,179) have a single metal, with smaller fractions containing two (16%, 352 out of 2,179) or three (< 1%, 11 out of 2,179) metals, with only a single MOF containing four metals. In cases where there are multiple metals in a single MOF (e.g. 364 out of 2,179), all metals are counted and included. Metals are ordered by their atomic number. Gray dotted lines indicate a change in the period of the periodic table.

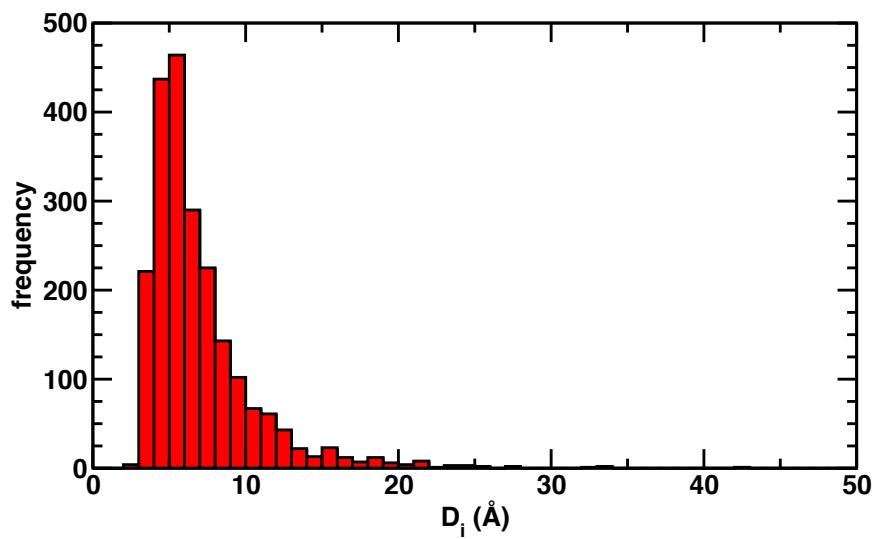

**Figure S7**. Distribution of pore sizes in the solvent-removal stability data set as quantified by the maximum included sphere ($D_i$) geometric descriptor, in angstroms. The maximum included sphere is the diameter of the largest sphere inside of the pores of the metal-organic framework. Larger-pore MOFs will have larger. $D_i$ values The bin size on the histogram is 1 Å. The smallest-pore MOF has a maximum included sphere of 2.9 Å, and the largest has a maximum included sphere of 42.8 Å, with a mean of 6.9 Å.

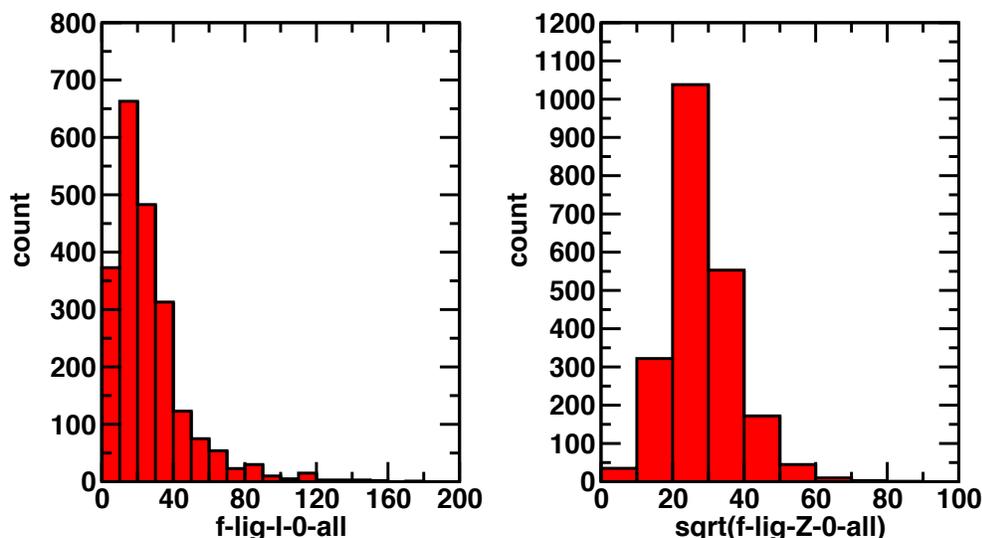

**Figure S8**. Distributions of two RAC properties of linkers in the solvent-removal stability dataset. The first property is the full-scope linker autocorrelation at depth 0 for the identity, which is equivalent to the number of atoms in the linker. This value ranges from two atoms to 172 atoms, demonstrating the diversity of linker size present in the dataset. Similarly, we plot the distribution of the square root of the full-scope linker autocorrelation at depth 0 for the nuclear charge, which is a proxy for the molecular weight (the RAC feature is proportional to the molecular weight). The bin width for both histograms is 10 units.

**Table S8**. Absolute counts of the number of stable and unstable MOFs in the 2,179 MOF solvent-removal stability dataset, as determined by natural language processing (NLP). Due to publishing bias (e.g. stable MOFs are more likely to be published than unstable MOFs), we observe a bias towards MOFs that are labeled "stable" upon solvent removal.

| label | count |
|---|---|
| stable (1) | 1,296 |
| unstable (0) | 883 |

**Table S9.** Table of the top twenty features correlated with the solvent-removal stability flag over the solvent removal stability dataset. Due to the binary nature of the solvent-removal stability label and the continuous nature of the input variables, we quantify the correlation using a point-biserial correlation coefficient, which is analogous to the Pearson correlation coefficient. We observe the features most highly correlated with solvent-removal stability to characterize the metal identity (e.g. of the metal-oxyhydroxy cluster), the linker coordinating atoms, the linker size, or the pore volume. No individual feature correlates strongly with the text-mined labels.

| variable | point-biserial correlation coefficient | interpretation of variable |
|---|---|---|
| mc-S-3-all | 0.16 | metal-centered depth 3 covalent radius product |
| $D_{mc}$-chi-3-all | -0.15 | metal-centered depth 3 electronegativity difference |
| lc-chi-0-all | 0.14 | linker-centered depth 0 electronegativity product |
| mc-T-3-all | 0.13 | metal-centered depth 3 topology product |
| mc-I-3-all | 0.12 | metal-centered depth 3 identity product |
| lc-Z-0-all | 0.12 | linker-centered depth 0 nuclear charge product |
| $D_i$ | 0.12 | maximum included sphere diameter (pore volume) |
| $D_f$ | 0.11 | maximum free sphere diameter |
| $D_{if}$ | 0.11 | maximum included sphere diameter in free sphere path |
| lc-Z-2-all | 0.11 | linker-centered depth 2 nuclear charge product |
| mc-T-2-all | 0.10 | metal-centered depth 2 topology product |
| $D_{lc}$-S-2-all | -0.10 | linker-centered depth 2 covalent radius difference |
| $D_{mc}$-chi-1-all | -0.10 | metal-centered depth 1 electronegativity difference |
| mc-chi-3-all | 0.10 | metal-centered depth 3 electronegativity product |
| $D_{mc}$-S-3-all | 0.10 | metal-centered depth 3 covalent radius difference |
| f-lig-Z-2-all | 0.10 | full-linker depth 2 nuclear charge product |
| $D_{mc}$-T-3-all | 0.10 | metal-centered depth 3 topology difference |
| f-lig-chi-0-all | 0.10 | full-linker depth 0 electronegativity product |
| f-lig-Z-0-all | 0.10 | full-linker depth 0 nuclear charge product (proxy for linker molecular weight) |
| f-lig-Z-1-all | 0.10 | full-linker depth 1 nuclear charge product |

**Table S10**. Performance of all machine learning (ML) models on the full feature set. Both the model accuracy and area under the receiver operating characteristic curve (AUC) are reported for each model, for the final training and test sets.

| solvent removal stability | | |
|---|---|---|
| model | train/val | test |
| SVM (RBF kernel) | accuracy: 1.00, AUC: 1.00 | accuracy: 0.74, AUC: 0.75 |
| GPC (RBF kernel) | accuracy: 0.96, AUC: 0.99 | accuracy: 0.73, AUC: 0.79 |
| GPC (Matern-3/2 kernel) | accuracy: 0.96, AUC: 1.00 | accuracy: 0.75, AUC: 0.79 |
| GPC (Matern-5/2 kernel) | accuracy: 0.96, AUC: 1.00 | accuracy: 0.75, AUC: 0.79 |
| ANN | accuracy: 1.00, AUC: 1.00 | accuracy: 0.76, AUC: 0.79 |

**Table S11**. Performance of all ML models after performing feature selection. Both the accuracy and area under the receiver operating characteristic curve (AUC) are reported for each model, for the final training and test sets.

| | solvent removal stability | |
|---|---|---|
| model | train | test |
| SVM (RBF kernel) | accuracy: 0.87, AUC: 1.00 | accuracy: 0.75, AUC: 0.77 |
| GPC (RBF kernel) | accuracy: 0.97, AUC: 0.99 | accuracy: 0.76, AUC: 0.81 |
| GPC (Matern-3/2 kernel) | accuracy: 0.97, AUC: 1.00 | accuracy: 0.74, AUC: 0.80 |
| GPC (Matern-5/2 kernel) | accuracy: 0.96, AUC: 0.99 | accuracy: 0.75, AUC: 0.80 |

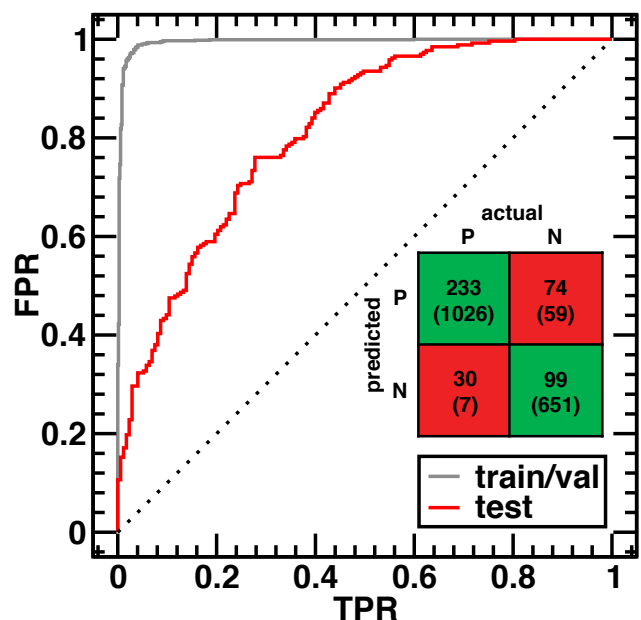

**Figure S9**. Receiver operating characteristic (ROC) curve for the feature-selected GPC model with an RBF kernel. In addition to the ROC curve for training (gray) and test (red) data, we include the confusion matrix for the test set, with the training set in parentheses. Due to dataset imbalance, the classifier has a slight bias towards predicting MOFs as "stable" hence having a greater number of false positives relative to false negatives. Rather than accuracy, which is sensitive to the number of data points in each class (e.g. the data balance), key quantities of the receiver operating characteristic (ROC) curve can indicate a more unbiased measure of classifier performance. The area under the curve (AUC) of the ROC curve can provide this measure, with a value of 0.5 indicating a random guess and a value of 1.0 indicating a perfect classifier.

**Table S12**. Table of the features selected during RF-RFA model training on solvent-removal stability. The scope of each feature is given in the table, and features are sorted by their scope (e.g. linker-only, metal-centered, then full unit cell). All autocorrelations starting with "D" are difference autocorrelations, and all others are product autocorrelations. The property that is autocorrelated (Z, T, I, S, $\chi$) is listed in the feature name, along with the corresponding depth.

| feature | scope |
|---|---|
| $D_{func}$-S-2-all | heteroatom functional group (linker only) |
| $D_{func}$-Z-1-all | heteroatom functional group (linker only) |
| $D_{func}$-$\chi$-1-all | heteroatom functional group (linker only) |
| func-S-0-all | heteroatom functional group (linker only) |
| func-S-1-all | heteroatom functional group (linker only) |
| func-Z-0-all | heteroatom functional group (linker only) |
| func-Z-1-all | heteroatom functional group (linker only) |
| func-Z-2-all | heteroatom functional group (linker only) |
| func-$\chi$-1-all | heteroatom functional group (linker only) |
| $D_{lc}$-S-1-all | linker coordinating atom centered (linker only) |
| $D_{lc}$-S-3-all | linker coordinating atom centered (linker only) |
| $D_{lc}$-T-2-all | linker coordinating atom centered (linker only) |
| $D_{lc}$-Z-1-all | linker coordinating atom centered (linker only) |
| $D_{lc}$-Z-2-all | linker coordinating atom centered (linker only) |
| $D_{lc}$-Z-3-all | linker coordinating atom centered (linker only) |
| lc-I-2-all | linker coordinating atom centered (linker only) |
| lc-I-3-all | linker coordinating atom centered (linker only) |
| lc-S-1-all | linker coordinating atom centered (linker only) |
| lc-S-2-all | linker coordinating atom centered (linker only) |
| lc-Z-1-all | linker coordinating atom centered (linker only) |
| lc-Z-2-all | linker coordinating atom centered (linker only) |
| lc-$\chi$-2-all | linker coordinating atom centered (linker only) |
| lc-$\chi$-3-all | linker coordinating atom centered (linker only) |
| f-lig-I-0-all | full linker (linker only) |
| f-lig-I-3-all | full linker (linker only) |
| f-lig-S-0-all | full linker (linker only) |
| f-lig-S-1-all | full linker (linker only) |
| f-lig-Z-0-all | full linker (linker only) |
| f-lig-Z-1-all | full linker (linker only) |
| f-lig-Z-2-all | full linker (linker only) |
| f-lig-Z-3-all | full linker (linker only) |
| f-lig-$\chi$-1-all | full linker (linker only) |
| f-lig-$\chi$-2-all | full linker (linker only) |
| f-lig-$\chi$-3-all | full linker (linker only) |
| $D_{mc}$-S-1-all | metal centered |
| $D_{mc}$-S-2-all | metal centered |
| $D_{mc}$-Z-2-all | metal centered |
| mc-$\chi$-3-all | metal centered |
| mc-Z-3-all | metal centered |
| mc-S-3-all | metal centered |
| mc-S-2-all | metal centered |
| mc-S-0-all | metal centered |
| mc-I-2-all | metal centered |
| f-I-0-all | full unit cell |
| f-S-0-all | full unit cell |
| f-Z-0-all | full unit cell |
| f-$\chi$-3-all | full unit cell |
| GSA | geometric |

**Figure S10**. Demonstration of binary classification using only linker and geometric features. This model performs worse (test accuracy: 64%, test AUC: 0.60) on unseen data than our GPC classifier. Although this decision tree incorporates features that would naively be expected to classify large-pore MOFs (e.g. $D_i$ and $D_{if}$), these features are not at the roots of the decision tree, where splits are determined by linker size and topologies. The conditions for each decision are given at each node, along with its Gini coefficient and the values categorized into true (left) and false (right). The value field shows the number of leaves in each branch of the tree.

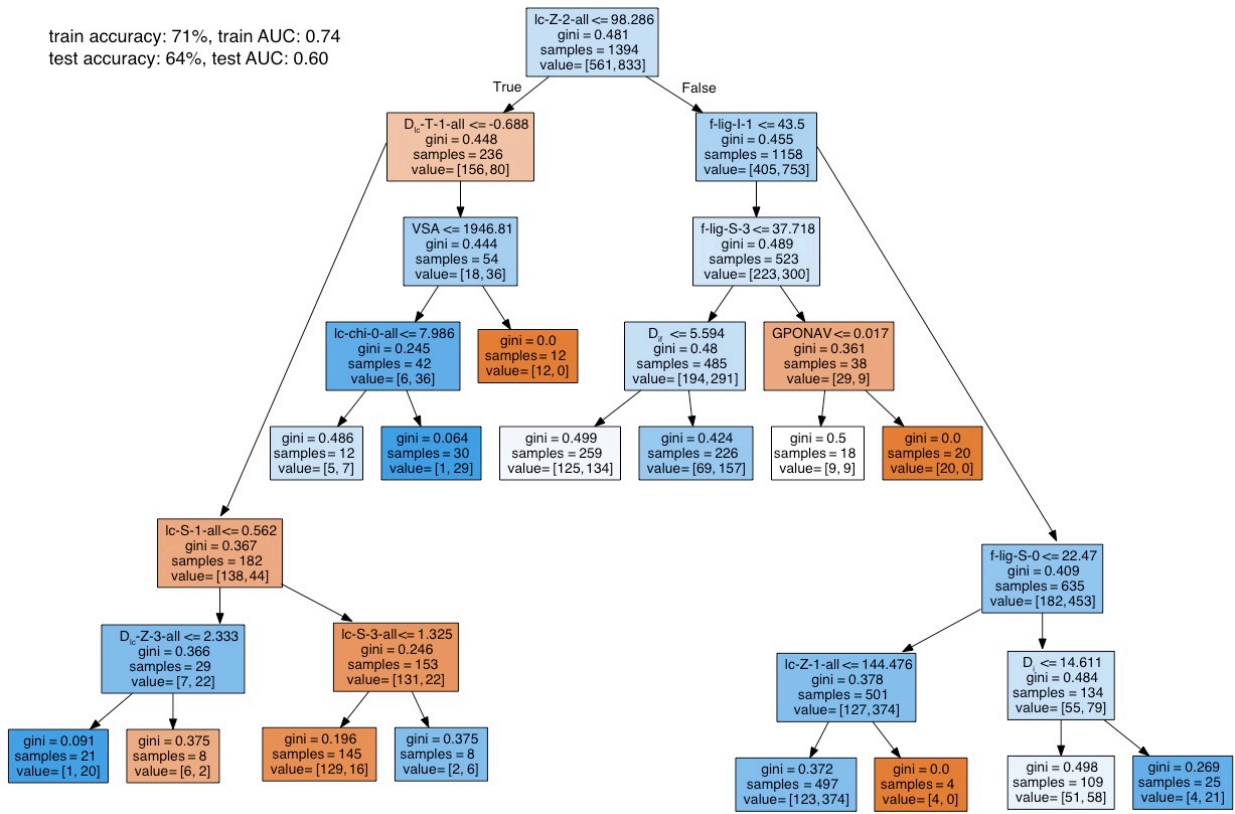

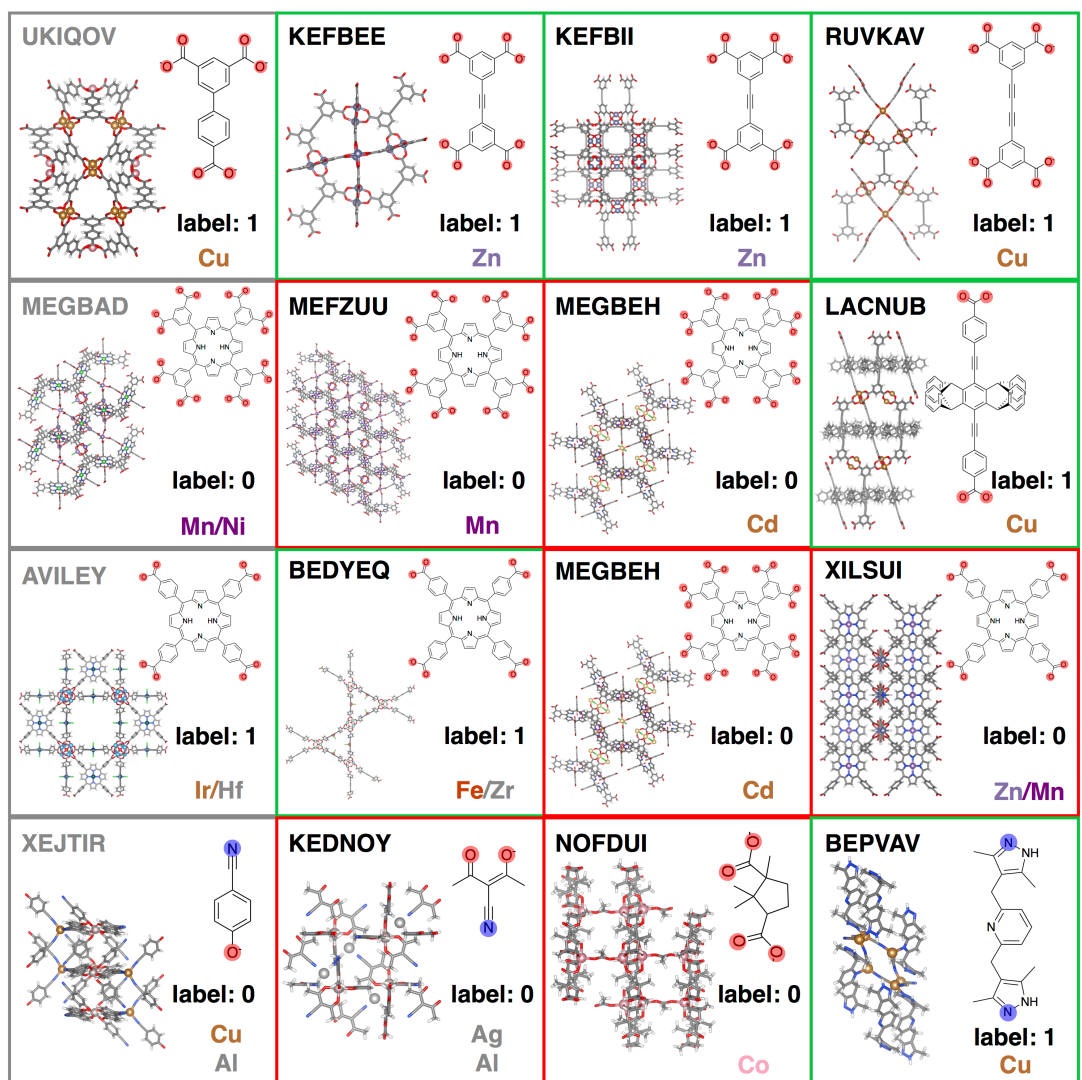

**Figure S11.** Four correctly predicted MOFs from the test set with their corresponding labels: stable (1) and unstable (0) with respect to solvent removal are shown in the left-hand column. The three nearest neighbors of each of these MOFs are shown, with increasing nearest-neighbor distance from left to right. UKIQOV (label: 1), MEGBAD (label: 0), AVILEY (label: 1), and XEJTIR (label: 0). The text-mined ground-truth values for the three nearest neighbors are colored on the diagram, with unstable in red and stable in green. Six digit identifiers are refcodes from the Cambridge Structural Database[1] (CSD). Within each MOF, the linkers and metals making up the MOF are shown, with the coordinating atoms colored by translucent circles, with oxygen in red and nitrogen in blue.

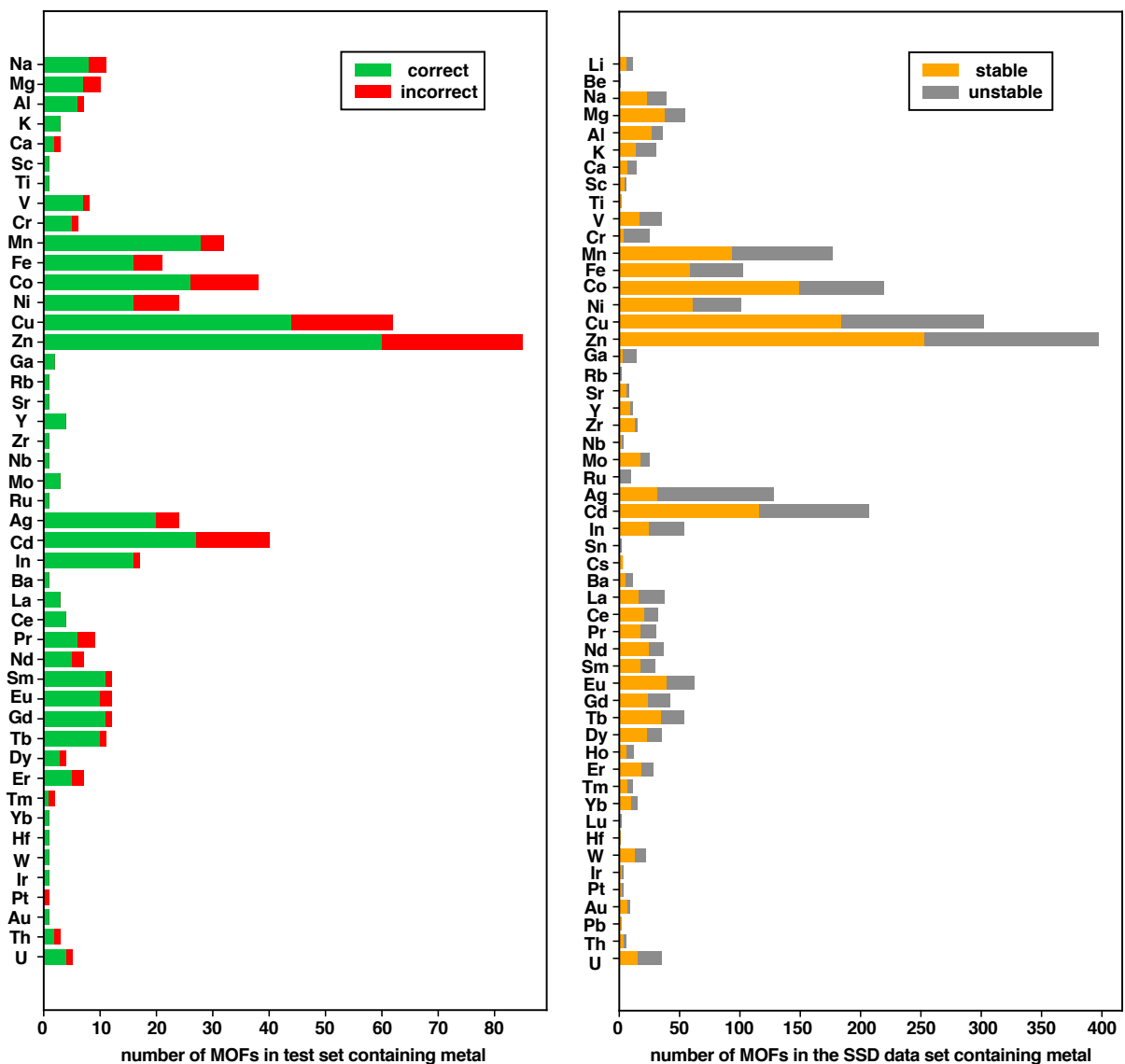

**Figure S12**. Unnormalized bar chart for prediction accuracy in the test set grouped by metal (left) and the overall distribution of stable and unstable MOFs in the solvent-removal stability data set grouped by metal (right). The classifier more accurately predicts MOFs with heavier metals (e.g. lanthanides), despite comparable fractions of unstable MOFs relative to the d-block transition metals. Metals that appear on the right pane but not the left pane are due to low frequencies (e.g. Lu), which result in the metal not being represented in the test set.

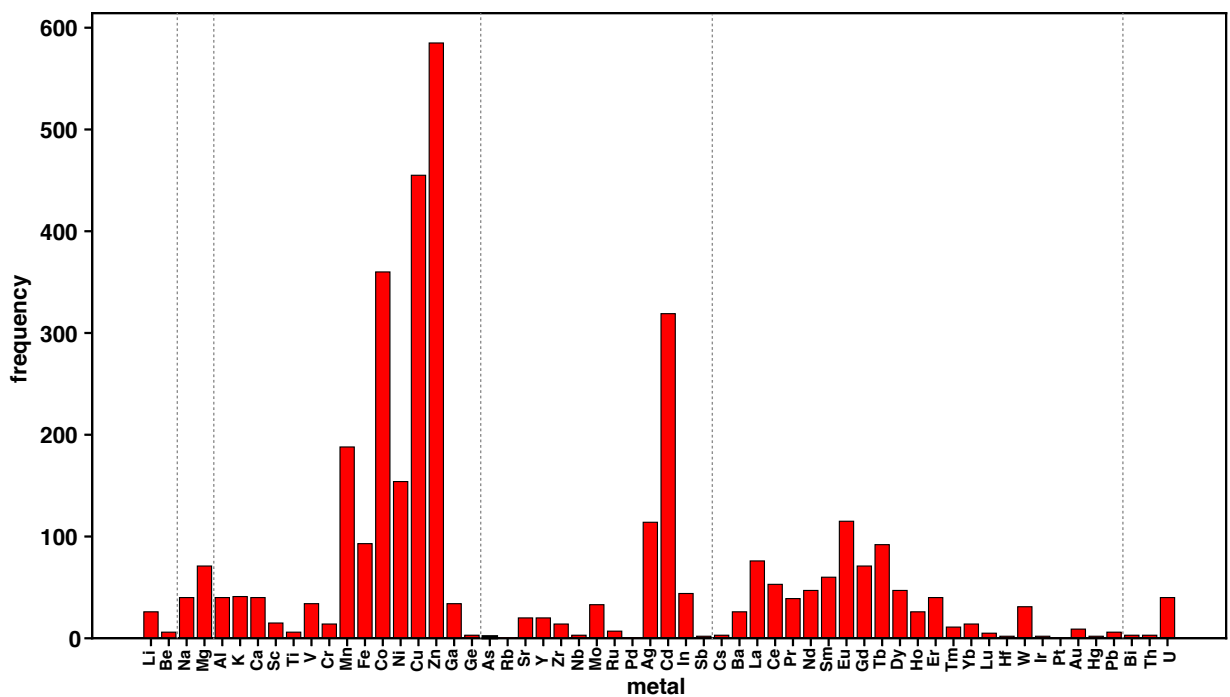

**Figure S13.** Frequencies of the metals that appear in the 3,132 MOFs in the thermal stability data set. The majority of MOFs (85%, 2,672 out of 3,132) have a single metal, with smaller fractions containing two (14%, 424 out of 3,132), three (< 1%, 26 out of 3,132), or four (< 1%, 9 out of 3,132) metals, with only a single MOF containing five metals. In cases where there are multiple metals in a single MOF (i.e., 460 out of 3,132), all metals are counted and included. Metals are ordered by their atomic number. Gray dotted lines indicate a change in the period of the periodic table.

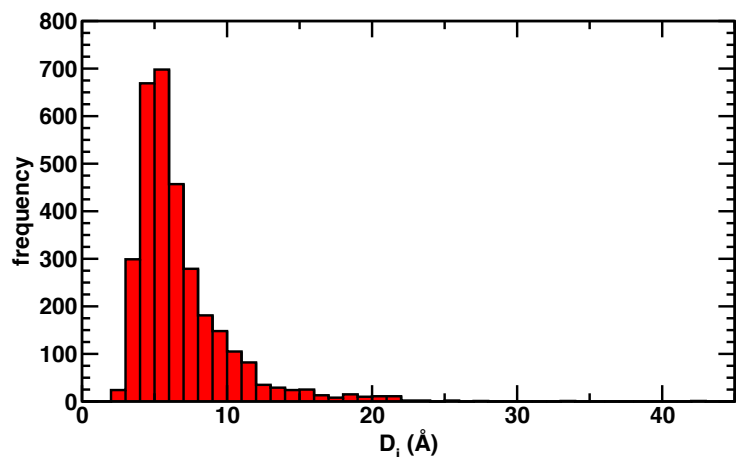

**Figure S14**. Distribution of pore sizes in the thermal stability data set as quantified by the maximum included sphere ($D_i$) geometric descriptor (in Å, bin width of 1 Å). The maximum included sphere is the diameter of the largest sphere inside of the pores of the metal-organic framework. The range in this set of $D_i$ is 2.8 Å to 42.8 Å, and the mean is 6.7 Å.

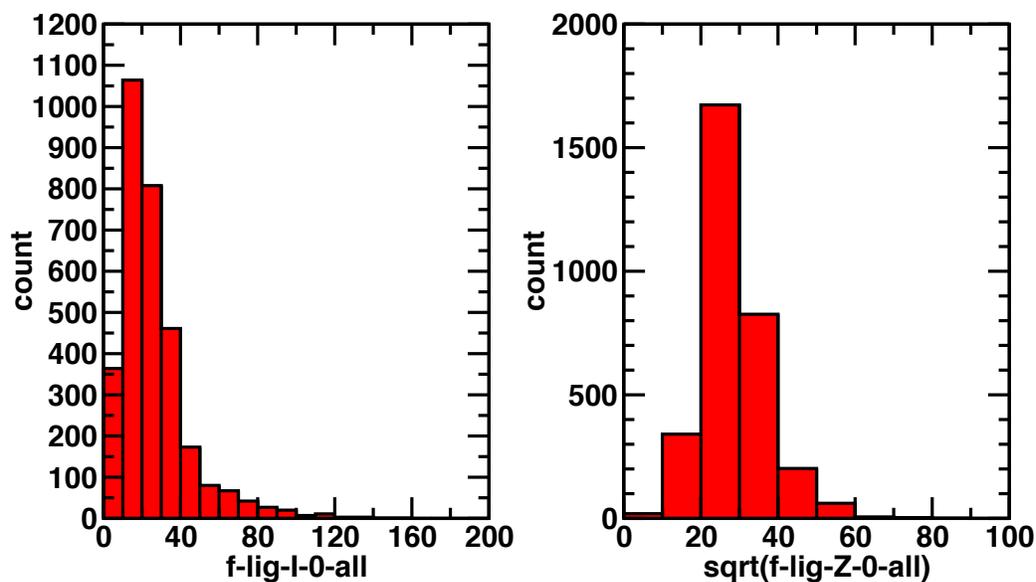

**Figure S15.** Distributions of two RAC properties of linkers in the thermal stability data set. The first property is the full-scope linker autocorrelation at depth 0 for the identity, analogous to the number of linker atoms and ranges from two to 156 atoms. The distribution of the square root of the full-scope linker autocorrelation at depth 0 for the nuclear charge, is analogous to the molecular weight. The bin width for both histograms is 10 units.

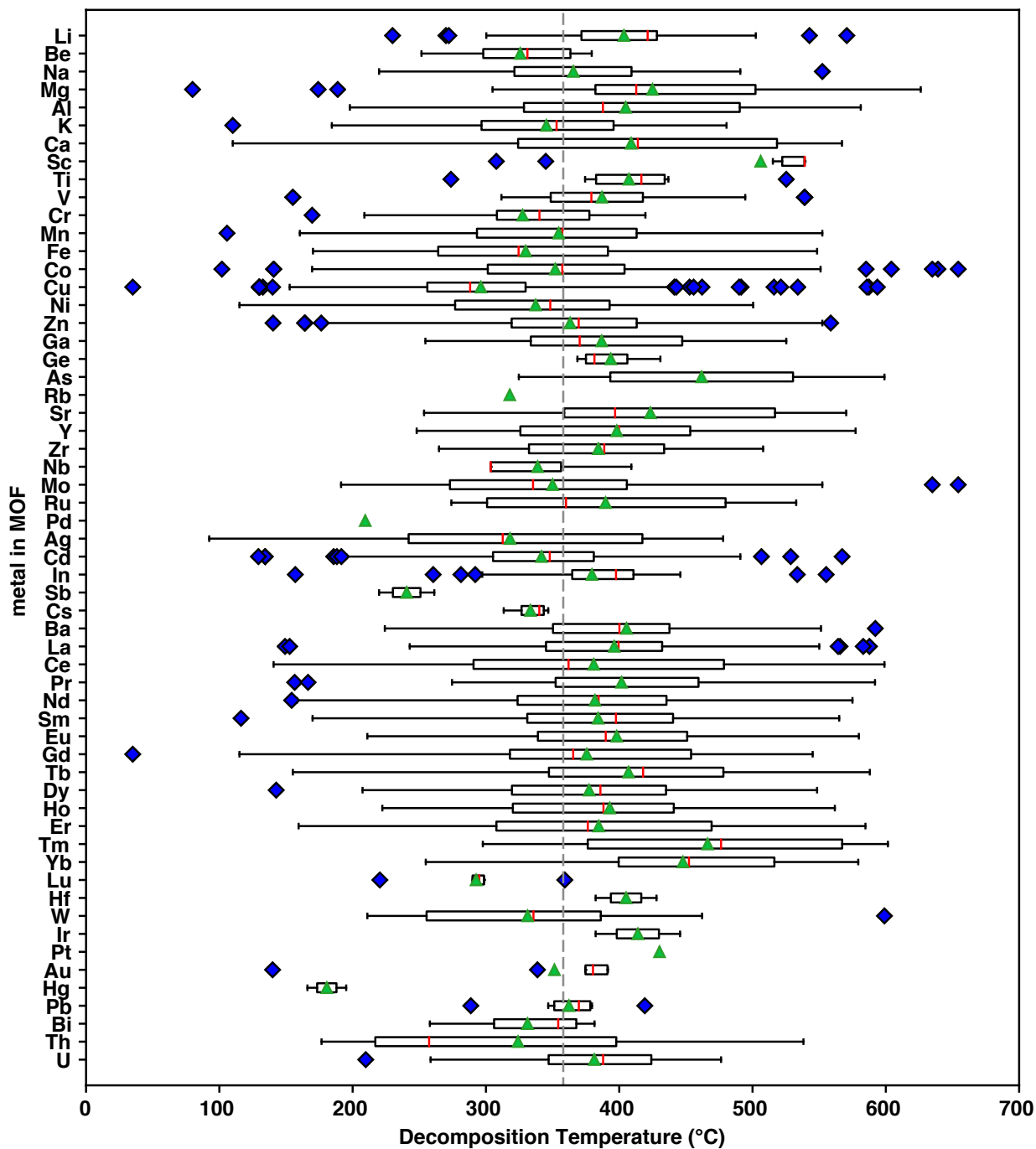

**Figure S16**. Boxplot of thermal decomposition temperatures by SBU metal (in °C) with red lines for the median and mean shown as a green triangle. The mean of the entire set is shown as a gray vertical dashed line. The box represents the first and third quartiles of data, and whiskers demonstrate bounds of 1.5× the interquartile range (quartile 3 − quartile 1), with any outliers shown as blue diamonds. If all data are within these bounds, the whiskers represent minima and maxima.

**Table S13.** Table of the top twenty features correlated (i.e., using the Pearson correlation coefficient) with the thermal decomposition temperature over the thermal stability dataset. The top features characterize the metal identity, the linker coordinating atoms, the linker size, or the pore volume. No individual feature correlates strongly with the TGA decomposition temperatures.

| variable | Pearson correlation coefficient | interpretation of variable |
|---|---|---|
| mc-chi-0-all | -0.34 | metal-centered depth 0 electronegativity product |
| $D_{mc}$-chi-1-all | -0.29 | metal-centered depth 1 electronegativity difference |
| $D_{mc}$-chi-3-all | -0.25 | metal-centered depth 3 electronegativity difference |
| $D_{mc}$-S-1-all | 0.24 | metal-centered depth 1 covalent radius difference |
| mc-S-0-all | 0.23 | metal-centered depth 0 covalent radius product |
| mc-chi-2-all | -0.23 | metal-centered depth 2 electronegativity product |
| mc-S-1-all | 0.21 | metal-centered depth 1 covalent radius product |
| $D_{lc}$-S-2-all | -0.21 | linker-centered depth 2 covalent radius difference |
| mc-I-1-all | 0.20 | metal-centered depth 1 identity product |
| $D_{lc}$-Z-2-all | -0.19 | linker-centered depth 2 nuclear charge difference |
| $D_{mc}$-T-1-all | 0.19 | metal-centered depth 1 topology difference |
| mc-T-0-all | 0.19 | metal-centered depth 0 topology product |
| $D_{lc}$-T-3-all | -0.19 | linker-centered depth 3 topology difference |
| $D_{mc}$-Z-1-all | 0.18 | metal-centered depth 1 nuclear charge difference |
| lc-chi-0-all | 0.17 | linker-centered depth 0 electronegativity product |
| mc-S-3-all | 0.17 | metal-centered depth 3 covalent radius product |
| mc-T-3-all | 0.17 | metal-centered depth 3 topology product |
| $D_{lc}$-chi-2-all | -0.17 | linker-centered depth 0 electronegativity difference |
| mc-T-1-all | 0.16 | metal-centered depth 1 topology product |
| mc-Z-3-all | 0.16 | metal-centered depth 3 nuclear charge product |

**Table S14**. Performance of all ML models on the TSD. For all models, the mean absolute error (MAE) and $R^2$ values are reported for the final training and test sets. Models without feature selection are shown first, followed by models with feature selection. We utilized kernel ridge regression (KRR), Gaussian process regression (GPR), and artificial neural network (ANN) models. All MAE values are reported in °C. Feature selection is not performed for the ANN model.

| thermal stability (no feature selection) | | |
|---|---|---|
| model | train errors | test errors |
| KRR (RBF kernel) | MAE: 26°C, $R^2$: 0.83 | MAE: 52°C, $R^2$: 0.39 |
| GPR (RBF kernel) | MAE: 15°C, $R^2$: 0.95 | MAE: 51°C, $R^2$: 0.38 |
| ANN | MAE: 13°C, $R^2$: 0.88 | MAE: 47°C, $R^2$: 0.39 |
| thermal stability (feature selection) | | |
| KRR (RBF kernel) | MAE: 19°C, $R^2$: 0.91 | MAE: 46°C, $R^2$: 0.43 |
| GPR (RBF kernel) | MAE: 12°C , $R^2$: 0.96 | MAE: 44°C, $R^2$: 0.46 |

**Table S15**. Table of the features selected during RF-RFA model training on thermal stability. The scope of each feature is listed in the table, and features are sorted by their scope (i.e., linker only, metal centered, then full unit cell). All features starting with "D" are difference autocorrelations, whereas the others are standard product-based autocorrelations. The property ($Z$, $T$, $I$, $S$, or $\chi$) is given in the feature name, along with the corresponding depth. Slightly more than half (26 out of 39) of the features characterize properties of the linker, with only 2 properties characterizing the unit cell, and the remaining 11 properties describing the SBU chemistry.

| feature | scope |
|---|---|
| func-Z-3-all | heteroatom functional group (linker only) |
| func-Z-2-all | heteroatom functional group (linker only) |
| func-T-3-all | heteroatom functional group (linker only) |
| func-T-0-all | heteroatom functional group (linker only) |
| func-$\chi$-1-all | heteroatom functional group (linker only) |
| func-I-0-all | heteroatom functional group (linker only) |
| $D_{func}$-Z-2-all | heteroatom functional group (linker only) |
| $D_{func}$-$\chi$-3-all | heteroatom functional group (linker only) |
| $D_{func}$-$\chi$-2-all | heteroatom functional group (linker only) |
| lc-$\chi$-2-all | linker coordinating atom centered (linker only) |
| lc-$\chi$-3-all | linker coordinating atom centered (linker only) |
| lc-T-3-all | linker coordinating atom centered (linker only) |
| $D_{lc}$-S-2-all | linker coordinating atom centered (linker only) |
| $D_{lc}$-Z-2-all | linker coordinating atom centered (linker only) |
| $D_{lc}$-$\chi$-2-all | linker coordinating atom centered (linker only) |
| f-lig-Z-1-all | full linker (linker only) |
| f-lig-S-1-all | full linker (linker only) |
| f-lig-$\chi$-0-all | full linker (linker only) |
| f-lig-S-2-all | full linker (linker only) |
| f-lig-$\chi$-1-all | full linker (linker only) |
| f-lig-I-1-all | full linker (linker only) |
| f-lig-Z-0-all | full linker (linker only) |
| f-lig-S-0-all | full linker (linker only) |
| f-lig-I-0-all | full linker (linker only) |
| f-lig-Z-2-all | full linker (linker only) |
| f-lig-T-0-all | full linker (linker only) |
| mc-$\chi$-0-all | metal centered |
| mc-$\chi$-2-all | metal centered |
| mc-Z-2-all | metal centered |
| mc-S-0-all | metal centered |
| mc-Z-0-all | metal centered |
| mc-$\chi$-1-all | metal centered |
| mc-T-1-all | metal centered |
| $D_{mc}$-Z-2-all | metal centered |
| $D_{mc}$-$\chi$-1-all | metal centered |
| $D_{mc}$-Z-1-all | metal centered |
| $D_{mc}$-S-2-all | metal centered |
| f-Z-2-all | full unit cell |
| f-Z-0-all | full unit cell |

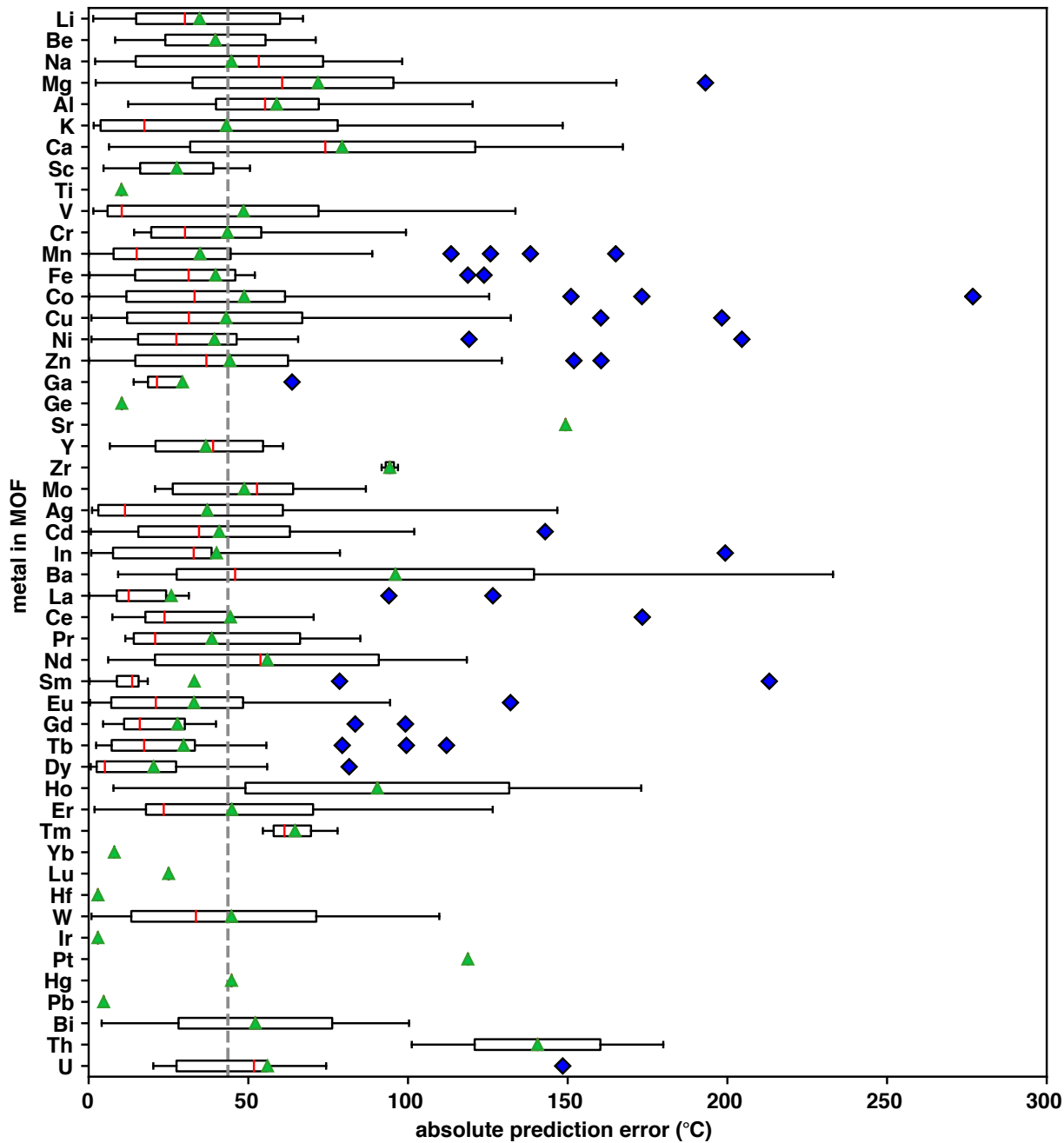

**Figure S17.** Boxplot of absolute prediction errors of thermal decomposition temperatures by metal (in °C) with red lines for the median and mean shown as a green triangle. The mean of the entire set is shown as a gray vertical dashed line. The box represents the first and third quartiles of data, and whiskers show bounds of 1.5× the interquartile range (quartile 3 − quartile 1), with any outliers shown as blue diamonds. If all data are within these bounds, the whiskers represent minima and maxima of the errors.

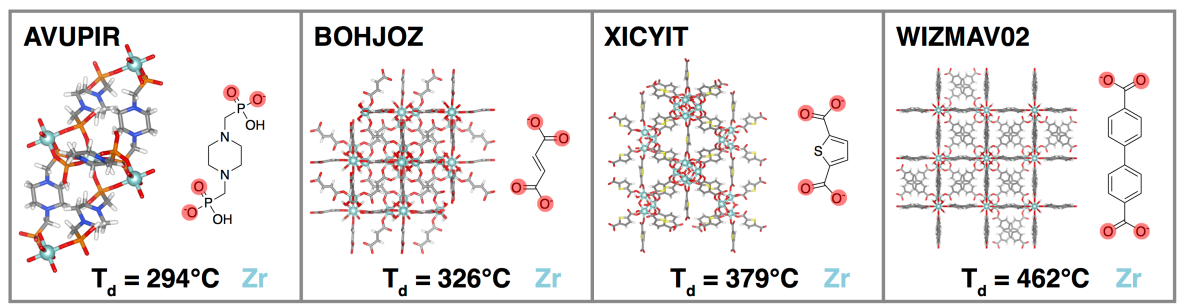

**Figure S18**. Four Zr MOFs from the thermal stability dataset and their corresponding thermal decomposition temperatures ($T_d$) shown inset. The MOFs are ordered from least to most thermally stable, with increasing $T_d$ from left to right with refcodes from the Cambridge Structural Database[1] (CSD) shown for all. Within each MOF, the linkers and metals making up the MOF are shown, with the coordinating oxygen atoms colored by translucent circles. Atoms in MOFs are colored as follows: white for hydrogen, gray for carbon, blue for nitrogen, red for oxygen, orange for phosphorus, yellow for sulfur, and light blue for zirconium. BOHJOZ is commonly known as MOF-801, XICYIT is commonly known as DUT-67, and WIZMAV02 is commonly known as UiO-67.

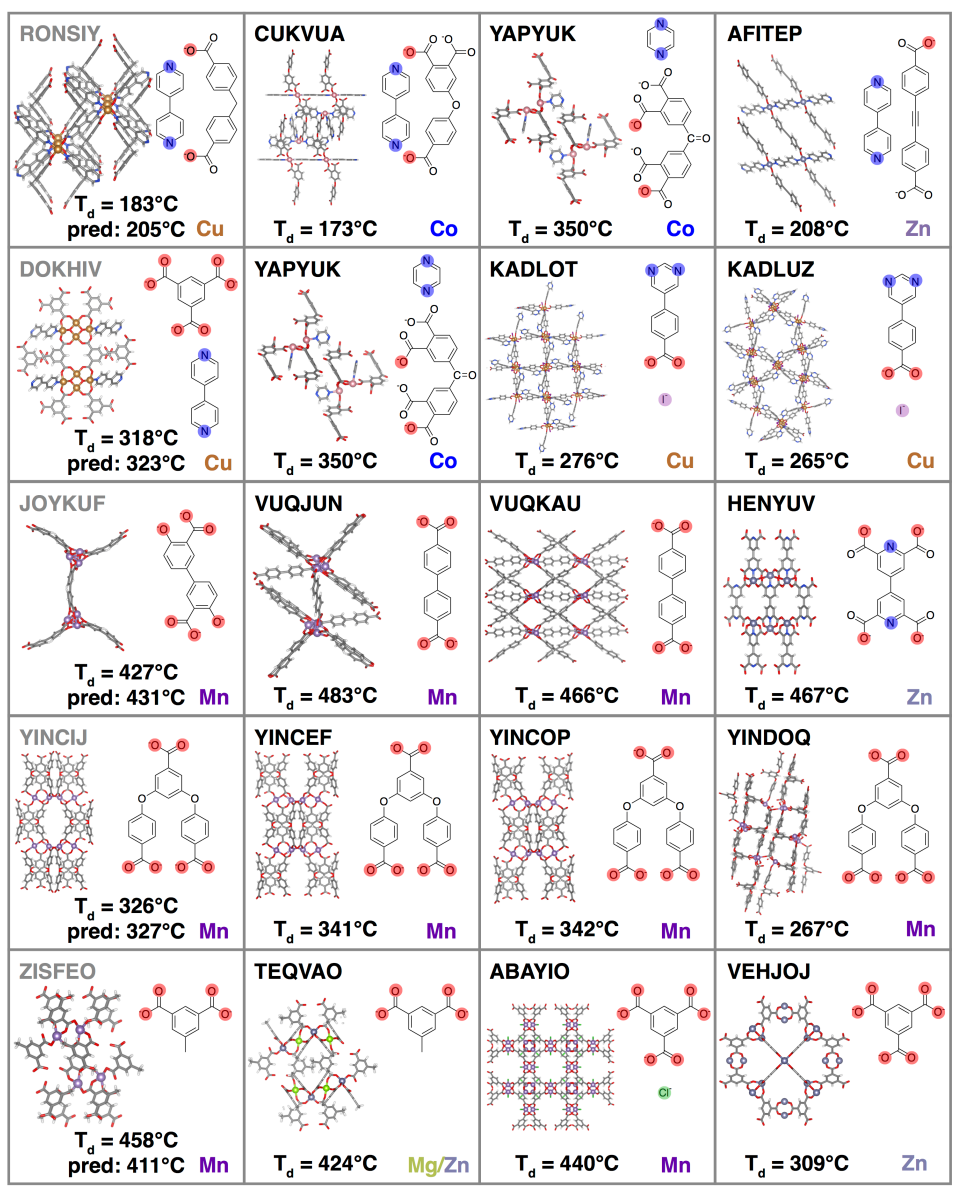

**Figure S19.** Five MOFs from the test set with low prediction errors and their corresponding thermal decomposition temperatures ($T_d$) shown on the left. GPR model predictions are shown in each box, labeled "pred". The three nearest neighbors for each of these MOFs are shown, with increasing nearest neighbor distance from left to right with refcodes from the Cambridge Structural Database[1] (CSD) shown for all. Within each MOF, the linkers and metals making up the MOF are shown, with the coordinating atoms colored by translucent circles, with oxygen in red, nitrogen in blue, chloride in green, and iodide in purple.

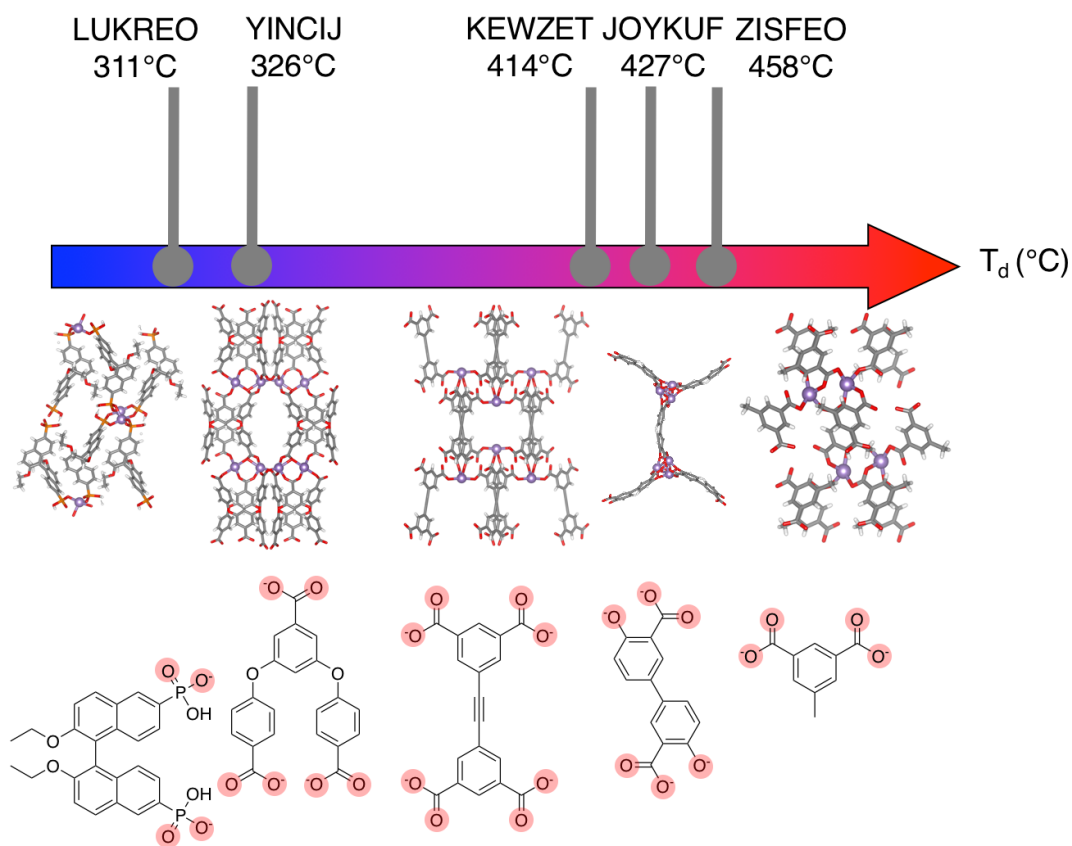

**Figure S20.** Examples of MOFs containing manganese SBUs with a range of linkers shown from left to right with increasing thermal decomposition temperatures ($T_d$, in °C). The MOF structure and linker are both shown with the linker's metal-coordinating atoms (here, all oxygen atoms) indicated as translucent circles in the skeleton structures. The linkers (left to right) are (2,2'-diethoxy-[1,1'-binaphthalene]-6,6'-diyl)bis(hydrogen phosphonate), 4,4'-((5-carboxylato-1,3-phenylene)bis(oxy))dibenzoate, 5,5'-(ethyne-1,2-diyl)diisophthalate, 4,4'-dioxido-[1,1'-biphenyl]-3,3'-dicarboxylate, and 5-methylisophthalate. The CSD refcodes for each structure are denoted above the $T_d$.

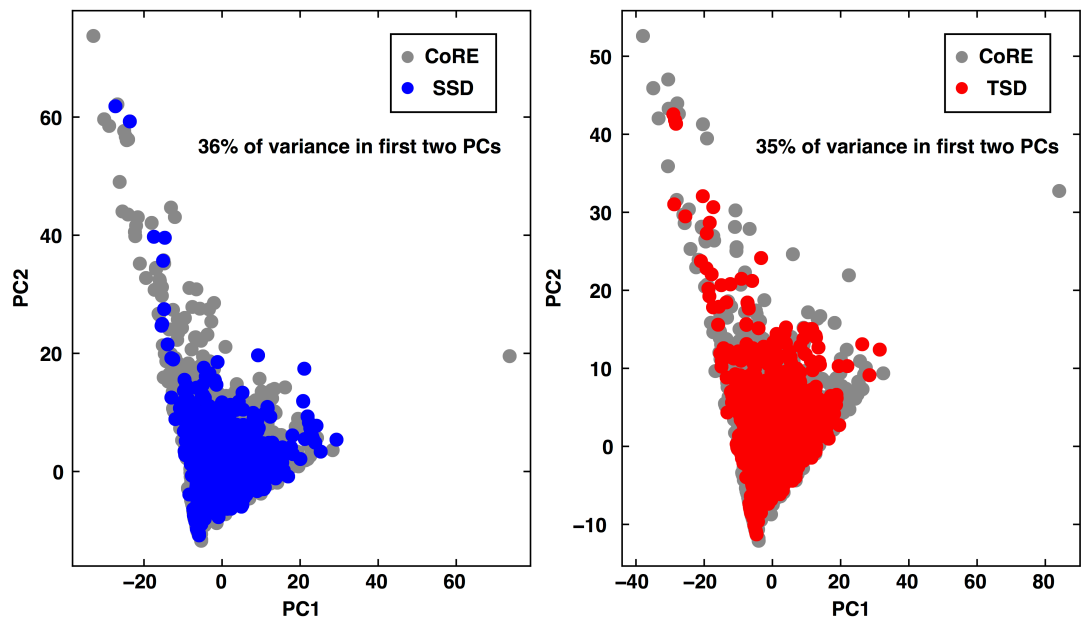

**Figure S21.** Principal component analyses of the solvent-removal stability data set (left, labeled SSD) and thermal stability data set (right, labeled TSD) relative to the full featurizable CoRE MOF data set (shown in gray on both). The first two principal components are shown, with the variance noted inset.

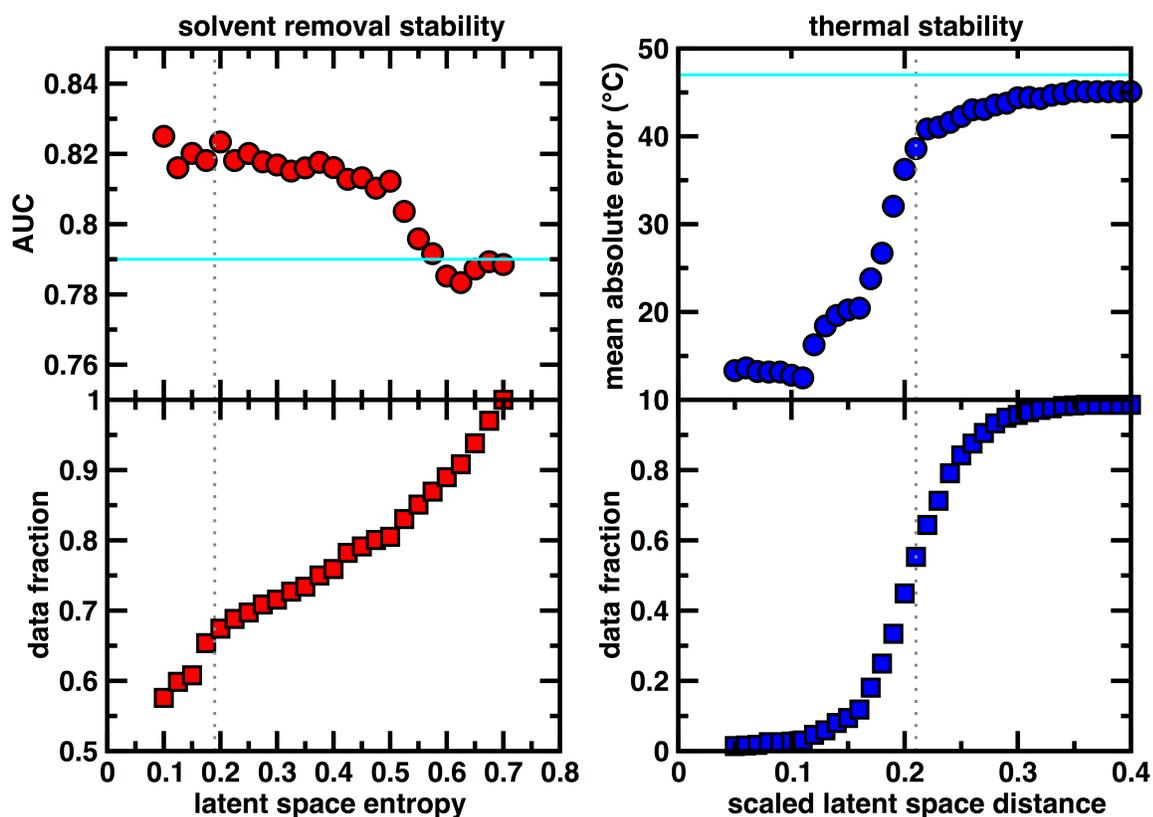

**Figure S22.** Latent space entropy (LSE) cutoff vs AUC and retained data fraction for the solvent-removal stability test data (left), and latent space distance (LSD) cutoff vs mean absolute error (MAE) and retained data fraction for the thermal stability test data (right). The most distant point in the thermal stability test data is scaled to have a latent distance of 1.0, but the x-axis range is then truncated to focus on the range of latent distance cutoffs that affect most of the data. The average LSE and LSD from the respective test data sets are shown as gray dotted lines. These are used as conservative uncertainty quantification (UQ) cutoffs to screen unseen data. Limiting predictions to this LSE cutoff of 0.19 on the solvent-removal stability test set leads to predictions on 66% of the test data, and an improved accuracy and AUC of 81% and 0.82 respectively. Similarly, limiting predictions to a scaled LSD cutoff of 0.21 on the thermal stability test set leads to predictions on 51% of the test data with a reduced mean absolute error (MAE) of 39°C. The test set AUC of 0.79 for solvent removal stability classification and test set MAE of 47°C are shown as cyan lines. We used 50 nearest neighbors to compute the latent space distances and entropies.

**Table S16.** Absolute counts of the number of MOFs from the featurizable CoRE MOF dataset (9,597 MOFs) where we cannot obtain an experimental result via automated text mining for solvent removal stability (left) or thermal stability (right).

| data set | solvent removal stability count | thermal stability count |
|---|---|---|
| featurizable CoRE MOF set | 9,597 | 9,597 |
| text mined experimental result | 2,179 | 3,132 |
| no automatically mined result | 7,418 | 6,465 |

**Table S17.** Absolute counts of the number of MOFs from the featurizable CoRE MOF dataset (9,597 MOFs) that do not overlap with training, validation, or test data of the solvent removal stability or thermal stability datasets and fall within the defined UQ cutoffs (solvent removal stability LSE cutoff: 0.19, thermal stability scaled LSD cutoff: 0.21). After eliminating data that appears in sets used for model training, 1,492 design space points remain for prediction.

| data set | solvent removal stability count | thermal stability count |
|---|---|---|
| featurizable CoRE MOF set | 9,597 | 9,597 |
| does not appear in any train, validation, or test sets used for model training | 5,844 | 5,844 |
| within latent space UQ cutoff | 3,492 | 2,522 |
| within latent space UQ cutoff for both models simultaneously | 1,492 | 1,492 |

**Table S18.** Analysis of the time necessary to predict the solvent-removal stability and thermal stability of a representative MOF (refcode: SODQOT). We report the profiled time for generating a prediction on this MOF. The first three steps of the process (e.g. generating descriptors from the primitive cell) must be done per structure and are parallelizable. The last two steps (e.g. normalizing data and making a prediction) can be done in one step for all structures. We demonstrate the approximate time necessary to make predictions on 1,500 compounds parallelized over 8 CPU cores.

| step | time (seconds) |
|---|---|
| getting primitive cell from unit cell | 0.1 |
| generating RAC descriptors from primitive cell | 4.2 |
| generating geometric descriptors using Zeo++ | 9.1 |
| loading ANN models into memory | 11.3 |
| normalizing data | 1.2 |
| making a prediction | 3.0 |
| **time estimate for predictions of 1,500 MOFs parallelized over 8 CPUs** | |
| (1,500 MOFs × (0.1+4.2+9.1 seconds)/8 CPUs)+(15.5 seconds) = 2,528 seconds = 42 minutes | 2,528 |

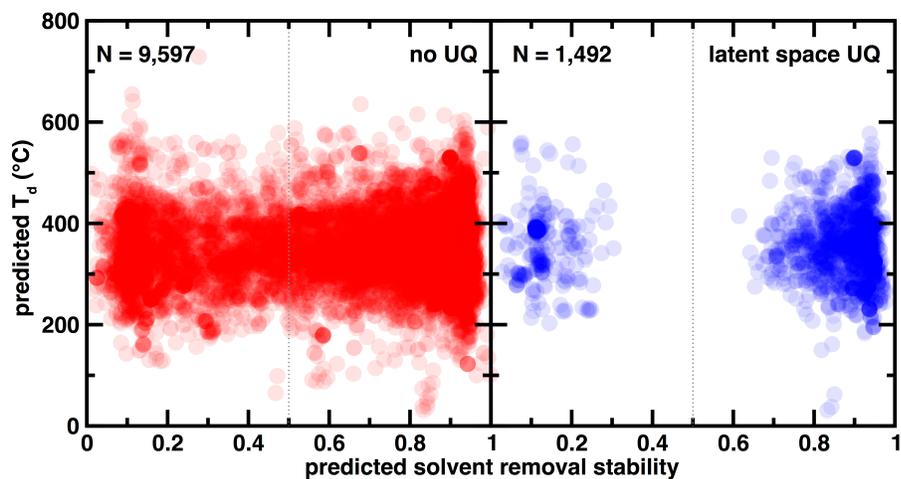

**Figure S23**. Application of our ML models over the non-UQ-constrained space of 9,597 CoRE MOFs (left) vs the UQ-constrained space of 1,492 MOFs (right). Data are colored with translucent circles to highlight data density. Employing strict UQ cutoffs over the design space restricted the model to only make predictions for data where it is more confident. An LSE cutoff of 0.19 and scaled LSD cutoff of 0.21 are used, because they are the average LSE and scaled LSD for the solvent-removal and thermal stability test sets respectively.

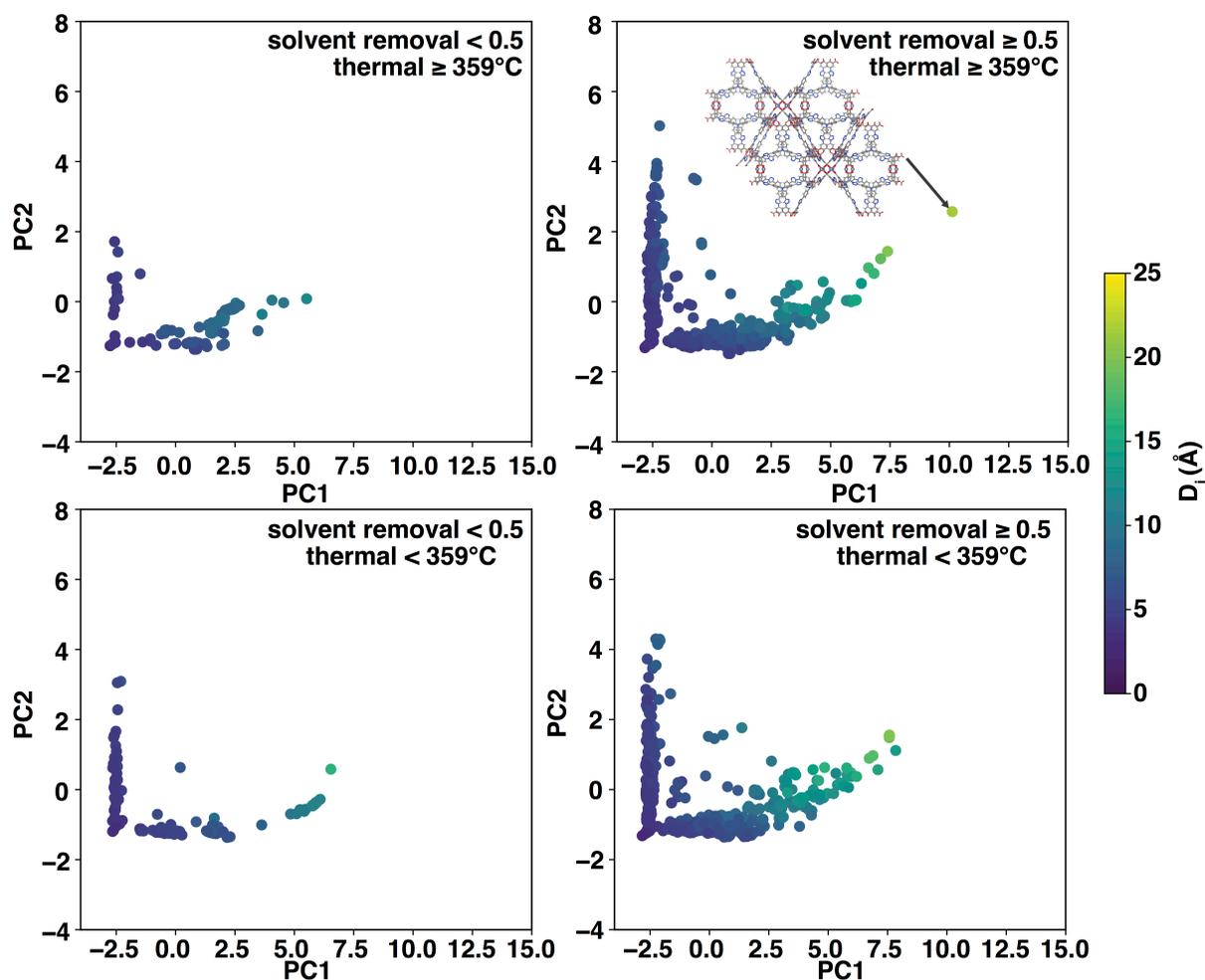

**Figure S24**. The first two components from a principal component analysis on the geometric features for the 1,492 unseen CoRE MOFs where we apply our models. MOFs are colored by their maximum included sphere value ($D_i$, in Å). Data are grouped by their stability classification: thermally stable but unstable with respect to solvent removal (top left), thermally stable and stable with respect to solvent removal (top right), thermally unstable and unstable with respect to solvent removal (bottom left) and thermally unstable but stable with respect to solvent removal (bottom right). The MOF with the largest maximum included sphere in this set (refcode: YUZPEQ, 22 Å) is highlighted on the top right, in the ball-and-stick representation. Zinc is shown in dark blue, oxygen in red, carbon in gray, nitrogen in blue, and hydrogen in white.

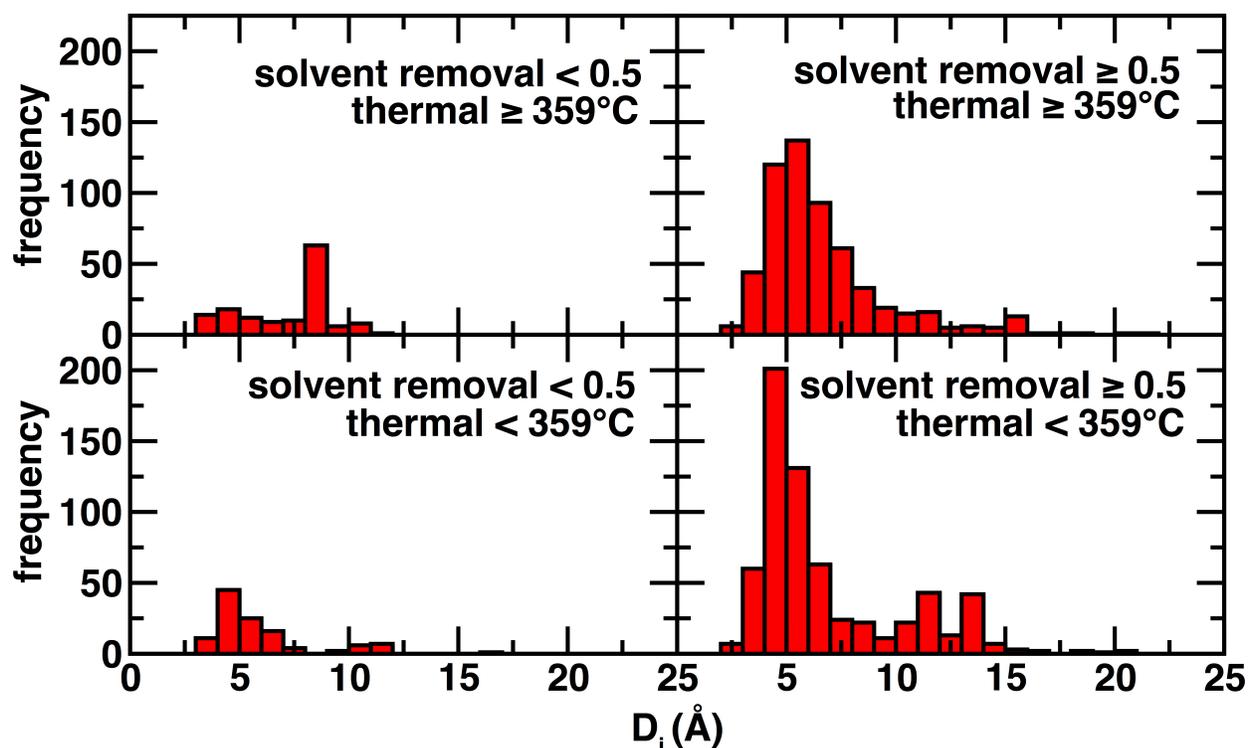

**Figure S25.** Histograms of the maximum included sphere ($D_i$, in Å) for the 1,492 unseen CoRE MOFs where we apply our models. Data are grouped by their stability classification: thermally stable but unstable with respect to solvent removal (top left), thermally stable and stable with respect to solvent removal (top right), thermally unstable and unstable with respect to solvent removal (bottom left) and thermally unstable but stable with respect to solvent removal (bottom right).

**Table S19.** Counts in the distinct quadrants of thermal and solvent removal stability as predicted by our ML models on the 1,492 MOFs falling within our UQ bounds. We define the quadrants by classifying MOFs into predictions of solvent removal stability and predictions of thermal decomposition temperature ($T_d$). Here, we characterize MOFs by their predicted solvent removal stability: < 0.5 (unstable) or ≥ 0.5 (stable), and thermal stability: $T_d$ < 359°C (less stable) or ≥ 359°C (more stable). Although $T_d$ is a continuous variable, we select a cutoff of 359°C, which is the average value of $T_d$ in the thermal stability data set.

| predicted solvent removal stability ≥ 0.5 | predicted thermal stability ≥ 359°C | counts |
|---|---|---|
| True | True | 578 |
| True | False | 141 |
| False | True | 656 |
| False | False | 117 |
| | | 1,492 |

**Figure S26**. Frequencies of metals appearing in the different quadrants of stability from the 1,492 MOFs falling within our UQ bounds. We define the quadrants by classifying MOFs into predictions of solvent removal stability and predictions of thermal decomposition temperature ($T_d$). Here, we characterize MOFs by their predicted solvent removal stability: < 0.5 (unstable) or ≥ 0.5 (stable), and thermal stability: $T_d$ < 359°C (less stable) or ≥ 359°C (more stable). Although $T_d$ is a continuous variable, we select a cutoff of 359°C, which is the average value of $T_d$ in the thermal stability data set. Counts of metals that appear in each quadrant are shown, with each subplot containing an inset label of its stability measure.

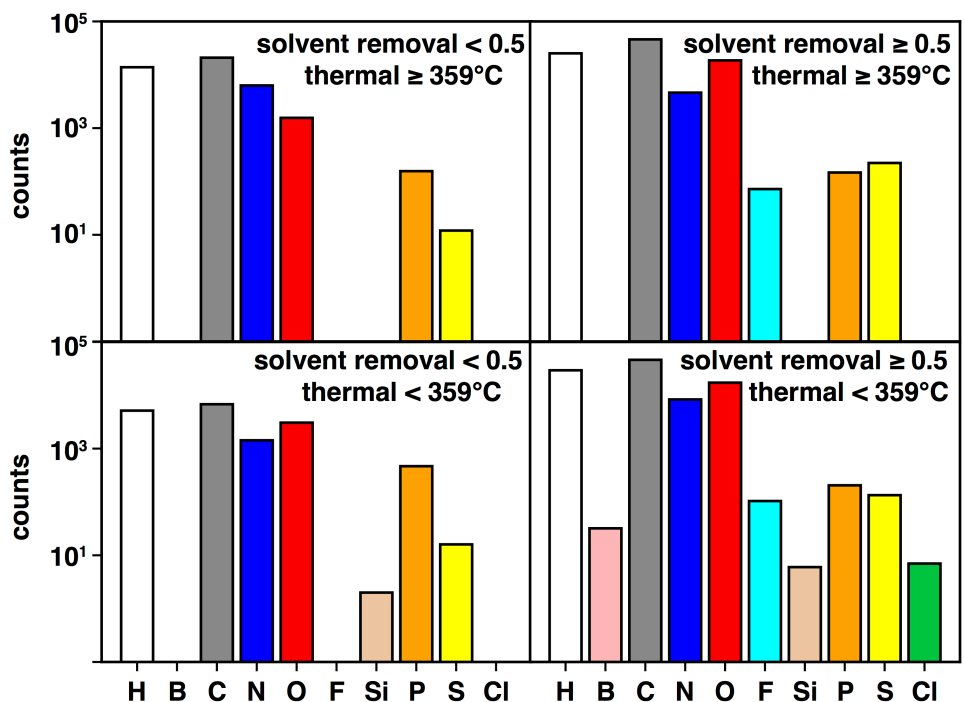

**Figure S27**. Atom counts for linkers in the 1,492 unseen CoRE MOFs where we apply our models. Elements are colored as follows: white for hydrogen, pink for boron, gray for carbon, blue for nitrogen, red for oxygen, cyan for fluorine, tan for silicon, orange for phosphorus, yellow for sulfur, and green for chloride. Data are grouped by their stability classification: thermally stable but unstable with respect to solvent removal (top left), thermally stable and stable with respect to solvent removal (top right), thermally unstable and unstable with respect to solvent removal (bottom left) and thermally unstable but high prevalence of hydrogen, carbon, and oxygen.

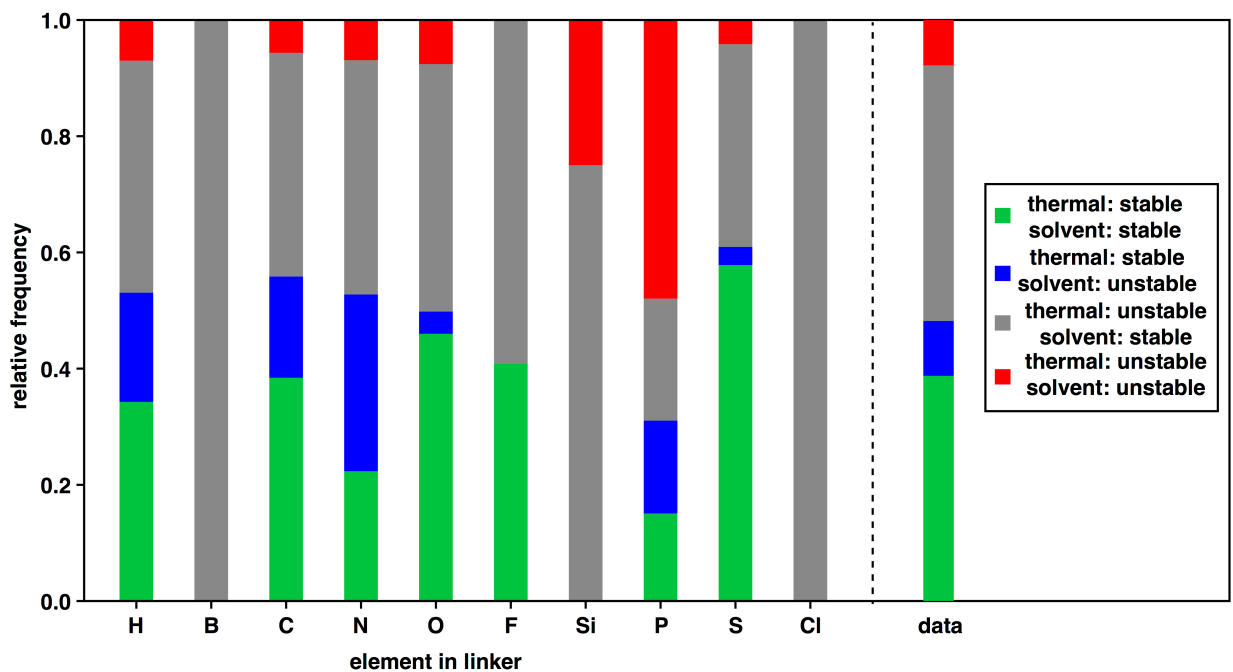

**Figure S28**. Atom-wise relative frequencies in a stacked bar chart based on the stability quadrants in which they appear. Elements are ordered on the x-axis by atomic mass and classified into the fraction that appear in each stability classification: thermally stable and stable with solvent removal (green), thermally stable but unstable with solvent removal (blue), thermally unstable but stable with solvent removal (gray), and thermally unstable and unstable with solvent removal (red). Because the number of MOFs that appear in each quadrant differs, a bar on the right shows the relative frequency of the number of MOFs in each set.

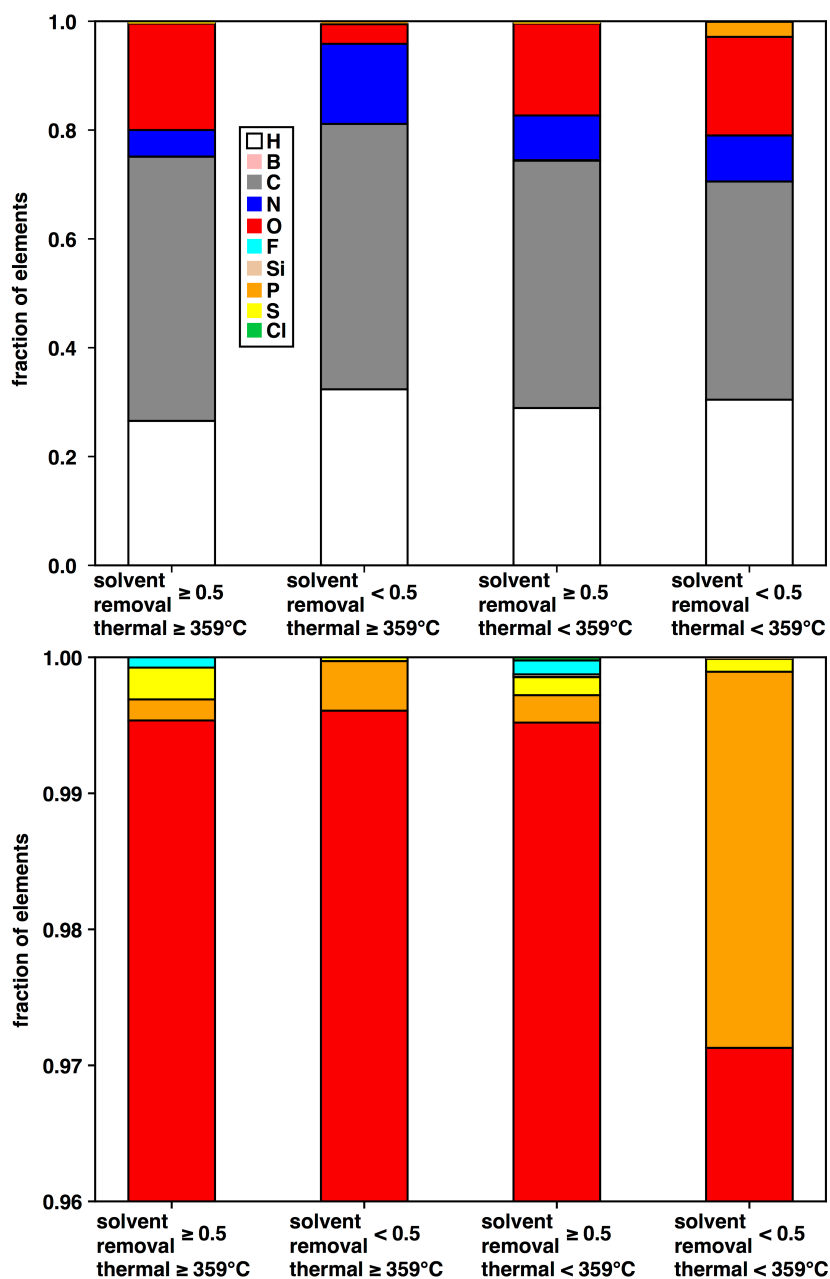

**Figure S29.** Relative frequency of atom counts for linkers in the 1,492 unseen CoRE MOFs where we apply our models. Elements are colored as follows: white for hydrogen, pink for boron, gray for carbon, blue for nitrogen, red for oxygen, cyan for fluorine, tan for silicon, orange for phosphorus, yellow for sulfur, and green for chloride. Data are grouped by their stability classification: thermally stable and stable upon solvent removal (left), thermally stable but unstable upon solvent removal (middle left), thermally unstable but stable upon solvent removal (middle right), and thermally unstable and unstable with solvent removal (right). A magnified pane (bottom) is shown above 0.96 to highlight less frequently occurring atoms.

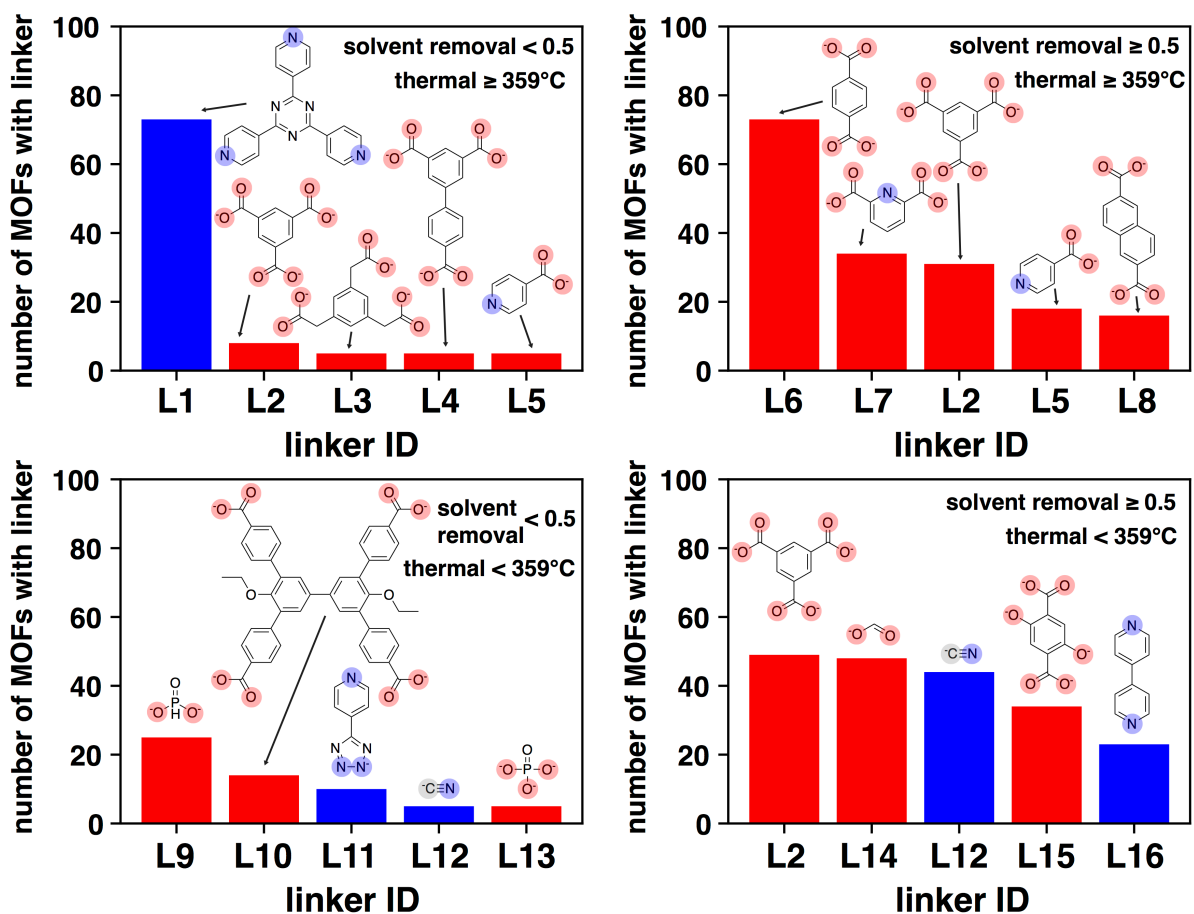

**Figure S30.** Frequency of the top five linkers in each quadrant of stability classification, colored by the heaviest metal-coordinating atom, with oxygen in red, and nitrogen in blue. Linkers are labeled by their linker IDs as follows: L1, 2,4,6-tri(pyridin-4-yl)-1,3,5-triazine; L2, benzene-1,3,5-tricarboxylate; L3, 2,2',2''-(benzene-1,3,5-triyl)triacetate; L4, [1,1'-biphenyl]-3,4',5-tricarboxylate; L5, isonicotinate; L6, terephthalate; L7, pyridine-2,6-dicarboxylate; L8, naphthalene-2,6-dicarboxylate; L9, phosphonate; L10, 5',5''-bis(4-carboxylatophenyl)-4'',6'-diethoxy-[1,1':3',1'':3'',1'''-quaterphenyl]-4,4'''-dicarboxylate; L11, 5-(pyridin-4-yl)tetrazol-2-ide; L12, cyanide; L13, phosphate; L14, formate; L15, dioxidoterephthalate; L16, 4,4'-bipyridine. Data are classified by their stability classification: thermally stable but unstable with respect to solvent removal (top left), thermally stable and stable with respect to solvent removal (top right), thermally unstable and unstable with respect to solvent removal (bottom left) and thermally unstable but stable with respect to solvent removal (bottom right).

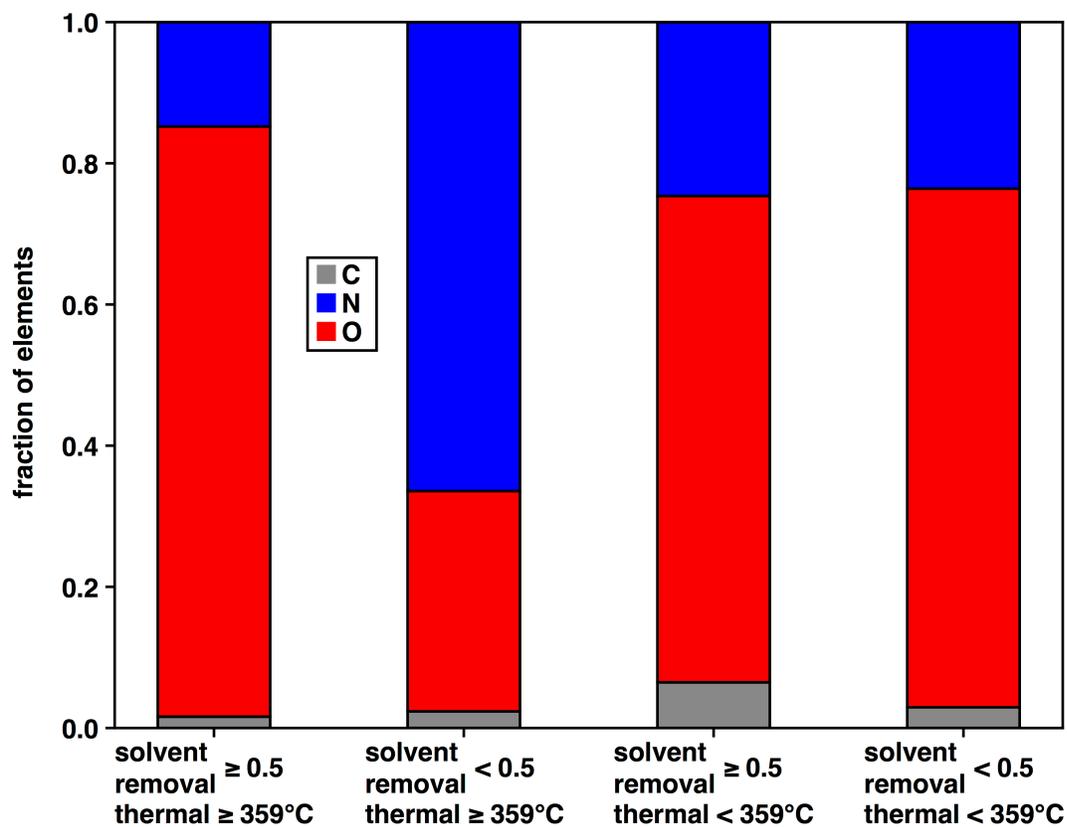

**Figure S31.** Relative frequency of linker connecting atom counts for linkers in the 1,492 unseen CoRE MOFs where we apply our models. Elements are colored as follows: gray for carbon, blue for nitrogen, and red for oxygen. Data are grouped by their stability classification: thermally stable and stable upon solvent removal (left), thermally stable but unstable upon solvent removal (middle left), thermally unstable but stable upon solvent removal (middle right), and thermally unstable and unstable with solvent removal (right).

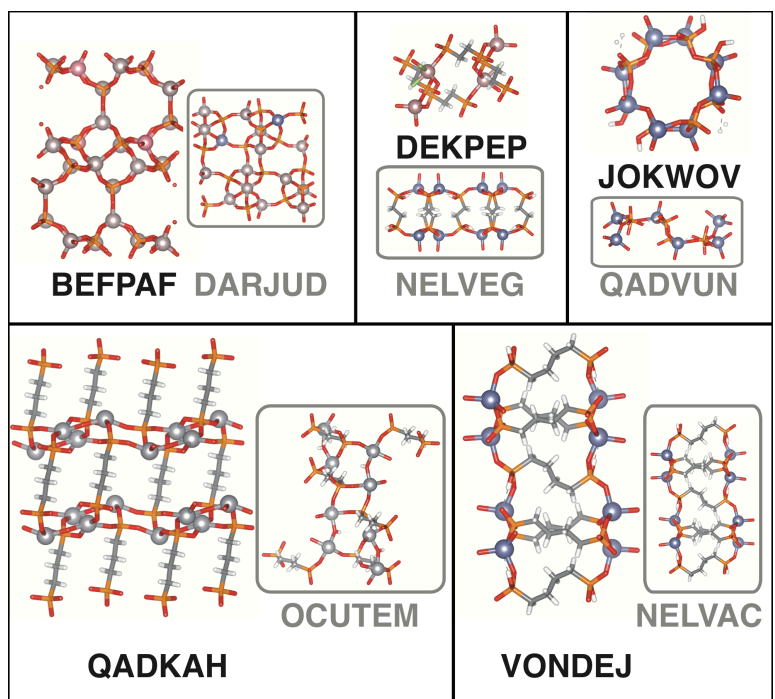

**Figure S32**. Example phosphonate MOFs identified by our models as MOFs that display both solvent-removal and thermal stability. Refcodes for each MOF are given below each structure in black. MOFs are shown in ball-and-stick representation, with hydrogen in white, carbon in gray, oxygen in red, fluorine in green, aluminum in silver, cobalt in pink, gallium in dark pink, zinc in dark blue, and vanadium in dark gray. Each MOF's nearest neighbor in the training data is shown with a gray outline and corresponding refcode.

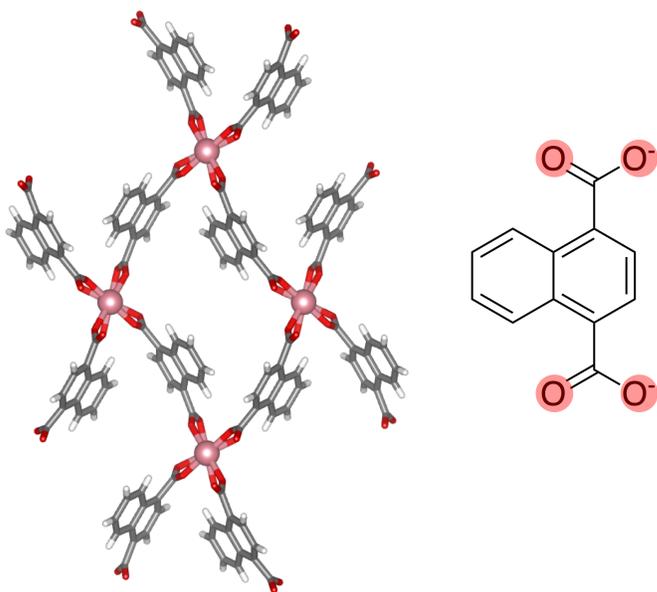

**Figure S33**. Example of a MOF (refcode: HEBTAL) predicted to be unstable with respect to solvent removal by our models. Analysis of the manuscript[2] shows that the MOF is indeed unstable with respect to solvent removal despite its small and rigid linker. Translucent red circles highlight the metal coordinating atoms on the linker (naphthalene-1,4-dicarboxylate).

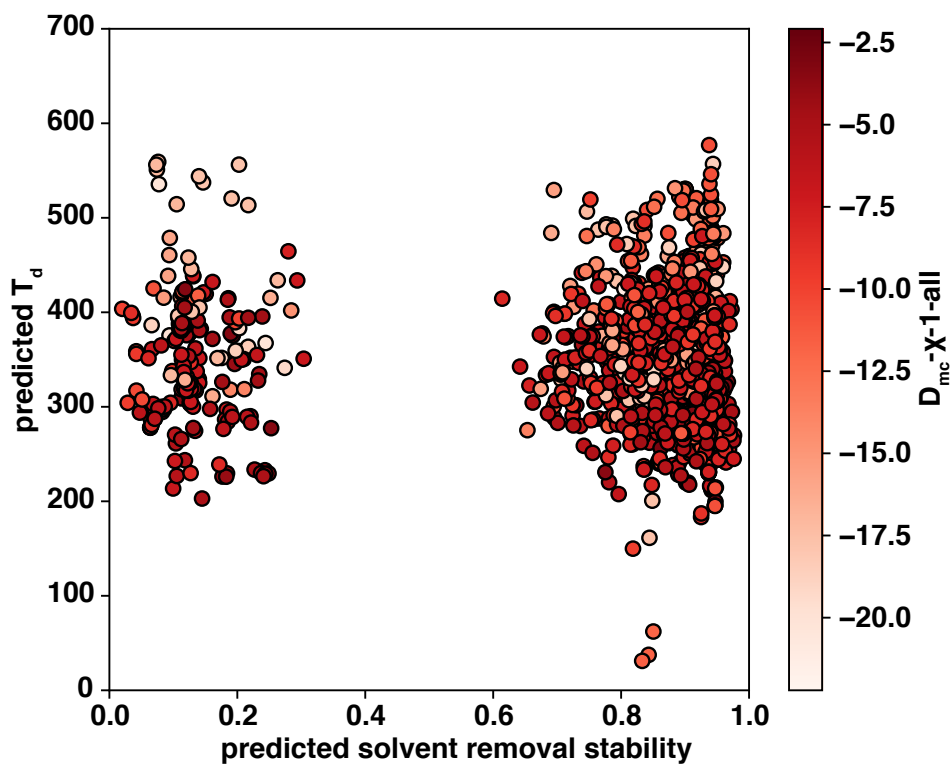

**Figure S34**. Predictions of solvent-removal stability and thermal stability over the 1,492 set of new MOFs, colored by their depth-1 metal centered electronegativity difference RAC feature. This feature encodes differences in electronegativity in the metal–linker bonds. Increasingly negative differences are indicated in white, and smaller electronegativity differences are indicated in red.

**Table S20**. The types of features and the number of descriptors for each scope are denoted. As in prior work, we employed a maximum depth of 3 for all RAC style descriptors, with five heuristic atom-wise quantities used to construct autocorrelations: nuclear charge (Z), topology (T), identity (I), electronegativity ($\chi$), and covalent radius (S). In addition to these RAC descriptors, 14 geometric descriptors are computed for each MOF: the maximum included sphere ($D_i$), maximum free sphere ($D_f$), maximum included sphere in the free sphere path ($D_{if}$), gravimetric pore volume (GPOV), volumetric pore volume (VPOV), gravimetric surface area (GSA), volumetric surface area (VSA), density ($\rho$), gravimetric pore accessible volume (GPOAV), gravimetric pore non-accessible volume (GPONAV), pore-accessible volume (POAV), pore non-accessible volume (PONAV), pore accessible void fraction (POAVF), and pore non-accessible void fraction (PONAVF).

| scope | RAC feature | maximum depth | quantities | number of descriptors |
|---|---|---|---|---|
| metal centered | y | 3 | Z, T, I, $\chi$, S | 20 product, 20 difference |
| linker connecting | y | 3 | Z, T, I, $\chi$, S | 20 product, 20 difference |
| functional group | y | 3 | Z, T, I, $\chi$, S | 20 product, 20 difference |
| full unit cell | y | 3 | Z, T, I, $\chi$, S | 20 product |
| full linker | y | 3 | Z, T, I, $\chi$, S | 20 product |
| geometry | n | N/A | $D_f$, $D_i$, $D_{if}$, GPOV, VPOV, GSA, VSA, $\rho$, POAV, PONAV, POAVF, PONAVF, GPOAV, GPONAV | 14 features |

**Table S21**. During model training, a subset of RACs are removed because they are constant. The final feature vector length is also denoted.

| features removed | final feature vector length |
|---|---|
| $D_{func}$-I-0-all, $D_{func}$-I-1-all, $D_{func}$-I-2-all, $D_{func}$-I-3-all, $D_{func}$-S-0-all, $D_{func}$-T-0-all, $D_{func}$-$\chi$-0-all, $D_{lc}$-I-0-all, $D_{lc}$-I-1-all, $D_{lc}$-I-2-all, $D_{lc}$-I-3-all, $D_{lc}$-S-0-all, $D_{lc}$-T-0-all, $D_{lc}$-Z-0-all, $D_{lc}$-$\chi$-0-all, $D_{mc}$-I-0-all, $D_{mc}$-I-1-all, $D_{mc}$-I-2-all, $D_{mc}$-I-3-all, $D_{mc}$-S-0-all, $D_{mc}$-T-0-all, $D_{mc}$-Z-0-all, $D_{mc}$-$\chi$-0-all, lc-I-0-all, mc-I-0-all | 148 (134 RACs + 14 geometric) |

**Table S22.** The hyperparameters used for various models are denoted in the table below. We report hyperparameters for models with and without feature selection (FS) for the gaussian process classifier (GPC), support vector machine (SVM) classifier, kernel ridge regression (KRR) regressor, and gaussian process regressor (GPR). In addition, we report the hyperparameters for an artificial neural network (ANN), for which we do not perform feature selection. For the SVM, two parameters exist: the kernel width ($\gamma$) and the regularization parameter (C). For the GPC and GPR models, the two hyperparameters are the kernel width ($\gamma$), and the noise level (variance). For GPC models, we additionally utilize the Matern-3/2 and Matern-5/2 kernels, which are generalizations of the RBF kernel.

| model | hyperparameters (no FS) | hyperparameters (FS) |
|---|---|---|
| solvent removal stability | | |
| SVM (RBF kernel) | C: 1977<br>$\gamma$: 0.09 | C: 4033<br>$\gamma$: 5.8 |
| GPC (RBF kernel) | variance: 1.557<br>$\gamma$: 3.231 | variance: 1.678<br>$\gamma$: 1.543 |
| GPC (Matern-3/2 kernel) | variance: 1.621<br>$\gamma$: 4.004 | variance: 1.783<br>$\gamma$: 1.751 |
| GPC (Matern-5/2 kernel) | variance: 1.592<br>$\gamma$: 3.695 | variance: 1.739<br>$\gamma$: 1.667 |
| ANN | architecture: (200, 200, 200)<br>learning rate: 0.00055088<br>dropout rate: 0.1119<br>batch size: 300<br>epochs: 1000<br>L2 regularization: 5.15e-5<br>patience for early stopping: 200<br>optimizer: Adam (with defaults) | N/A |
| thermal stability | | |
| KRR (RBF kernel) | C: 0.302<br>$\gamma$: 0.0229 | C: 0.336<br>$\gamma$: 0.839 |
| GPR (RBF kernel) | variance: 0.914<br>$\gamma$: 2.777 | variance: 0.888<br>$\gamma$: 1.103 |
| ANN | architecture: (200, 200)<br>learning rate: 0.0008815<br>dropout rate: 0.0321<br>batch size: 300<br>epochs: 4000<br>L2 regularization: 0.001<br>patience for early stopping: 200<br>optimizer: Adam (with defaults) | N/A |